\documentclass[twocolumn]{aastex7}

\usepackage{rotating}
\usepackage{threeparttablex}

\graphicspath{{./}{figures/}}

\renewcommand{\textbf}[1]{#1}

\begin{document}

\title{Shock properties for solar energetic particle events with signatures of inverse velocity arrival}

\author[orcid=0000-0001-6589-4509]{A. Kouloumvakos}
\affiliation{The Johns Hopkins University Applied Physics Laboratory, 11101 Johns Hopkins Road, Laurel, MD 20723, USA}
\email[show]{Athanasios.Kouloumvakos@jhuapl.edu}

\author[orcid=0000-0002-3176-8704]{D. Lario}
\affiliation{Heliophysics Science Division, NASA Goddard Space Flight Center, Greenbelt, MD 20771, USA.}
\email{david.larioloyo@nasa.gov}

\author[orcid=0000-0003-2169-9618]{G. M. Mason}
\affiliation{The Johns Hopkins University Applied Physics Laboratory, 11101 Johns Hopkins Road, Laurel, MD 20723, USA}
\email{Glenn.Mason@jhuapl.edu}

\author[orcid=0000-0002-8164-5948]{A. Vourlidas}
\affiliation{The Johns Hopkins University Applied Physics Laboratory, 11101 Johns Hopkins Road, Laurel, MD 20723, USA}
\email{Angelos.Vourlidas@jhuapl.edu}

\author[orcid=0000-0003-2079-5683]{R. C. Allen}
\affiliation{Southwest Research Institute, 6220 Culebra Rd. San Antonio, TX 78238, USA.}
\email{robert.allen@swri.org}

\author[orcid=0000-0001-6344-6956]{N. Wijsen}
\affiliation{Centre for mathematical Plasma Astrophysics, KU Leuven Campus Kulak, 8500 Kortrijk, Belgium}
\email{nicolas.wijsen@gmail.com}

\author[orcid=0000-0003-2865-1772]{X. Chen}
\affiliation{University of Michigan, 500 S. State Street, Ann Arbor, MI 48109, USA.}
\email{xiaohang1203@arizona.edu}

\author[orcid=0000-0002-9829-3811]{Z. Ding}
\affiliation{Institute of Experimental and Applied Physics, Christian-ALbrechts-University Kiel, 24118 Kiel, Germany}
\email{zheyiding96@gmail.com}

\author[orcid=0000-0001-6344-6956]{I. C. Jebaraj}
\affiliation{Department of Physics and Astronomy, University of Turku, FI-20500 Turku, Finland}
\email{immanuel.jebaraj@utu.fi}

\author[orcid=0000-0002-1859-456X]{P. Riley}
\affiliation{Predictive Science Inc. (PSI), 9990 Mesa Rim Road, Suite 170, San Diego, CA 92121, USA.}
\email{pete@predsci.com}

\author[0000-0001-6160-1158]{D. J. McComas}
\affiliation{Department of Astrophysical Sciences, Princeton University, Princeton, NJ 08544, USA}
\email{dmccomas@princeton.edu}

\author[orcid=0000-0002-0978-8127]{C. M. S. Cohen}
\affiliation{California Institute of Technology, 1260 E California Blvd, Pasadena, CA 91125, USA.}
\email{cohen@srl.caltech.edu}

\author[orcid=0000-0002-8387-5202]{E. Paouris}
\affiliation{The Johns Hopkins University Applied Physics Laboratory, 11101 Johns Hopkins Road, Laurel, MD 20723, USA}
\email{Evangelos.Paouris@jhuapl.edu}

\author[orcid=0000-0002-4381-3197]{S. Raptis}
\affiliation{The Johns Hopkins University Applied Physics Laboratory, 11101 Johns Hopkins Road, Laurel, MD 20723, USA}
\email{Savvas.Raptis@extcloud.jhuapl.edu}

\author[orcid=0000-0003-2361-5510]{L. Rodríguez-García}
\affiliation{European Space Agency (ESA), European Space Astronomy Centre (ESAC), Camino Bajo del Castillo s/n, 28692 Villanueva de la Cañada, Madrid, Spain}
\affiliation{Universidad de Alcal\'a, Space Research Group, 28805 Alcal\'a de Henares, Spain}
\email{Laura.RodriguezGarcia@esa.int}

\author[orcid=0000-0002-9246-996X]{Z. G. Xu}
\affiliation{California Institute of Technology, 1260 E California Blvd, Pasadena, CA 91125, USA.}
\email{zgxu@caltech.edu}

\author[orcid=0000-0001-6010-6374]{G. D. Berland}
\affiliation{The Johns Hopkins University Applied Physics Laboratory, 11101 Johns Hopkins Road, Laurel, MD 20723, USA}
\email{Grant.Berland@jhuapl.edu}

\author[orcid=0000-0003-1093-2066]{G. C. Ho}
\affiliation{Southwest Research Institute, 6220 Culebra Rd. San Antonio, TX 78238, USA.}
\email{george.ho@swri.org}

\author[orcid=0000-0003-1960-2119]{D. G. Mitchell}
\affiliation{The Johns Hopkins University Applied Physics Laboratory, 11101 Johns Hopkins Road, Laurel, MD 20723, USA}
\email{Donald.G.Mitchell@jhuapl.edu}

\author[orcid=0000-0002-2270-0652]{E. C. Roelof}
\affiliation{The Johns Hopkins University Applied Physics Laboratory, 11101 Johns Hopkins Road, Laurel, MD 20723, USA}
\email{Edmond.Roelof@jhuapl.edu}

\author[orcid=0000-0002-4240-1115]{J. Rodriguez-Pacheco}
\affiliation{Universidad de Alcal\'a, Space Research Group, 28805 Alcal\'a de Henares, Spain}
\email{javier.pacheco@uah.es}

\author[orcid=0000-0002-5674-4936]{M. E. Hill}
\affiliation{The Johns Hopkins University Applied Physics Laboratory, 11101 Johns Hopkins Road, Laurel, MD 20723, USA}
\email{Matthew.Hill@jhuapl.edu}

\author[orcid=0000-0002-7388-173X]{R. F. Wimmer-Schweingruber}
\affiliation{Institute of Experimental and Applied Physics, Christian-ALbrechts-University Kiel, 24118 Kiel, Germany}
\email{wimmer@physik.uni-kiel.de}

\begin{abstract}

We present a detailed investigation of the shock properties associated with solar energetic particle (SEP) events that exhibit a concave (``nose-like'') shape in their energy spectrogram, characterized by inverse velocity arrival (IVA) of the particles, where high-energy particles arrive later than mid-energy ones. Using measurements from Solar Orbiter and Parker Solar Probe between 2018 and 2025, we identify 26 such SEP events and reconstruct the observed shock fronts in three dimensions. We derive shock parameters along the magnetic field lines connected to each spacecraft using kinematic modeling and coronal magnetohydrodynamic simulations. Our analysis indicates that IVA-SEP events arise due to the spatial and temporal evolution of the shock properties and magnetic connectivity. In most cases analyzed here, the magnetic connectivity starts on the flanks of CME-driven shocks, where shocks tend to be weak, and shifts toward the shock apex sampling stronger portions of the shock front. This evolution of the shock properties at the connected field lines likely leads to the delayed arrival of high-energy particles and the progressive hardening of the SEP energy spectrum, observed in some of the events. We find a correlation between the transition energy at which the IVA begins and the shock speed along the connected field lines, consistent with expectations from time-dependent diffusive shock acceleration. Our results underscore the importance of the evolving shock properties, magnetic connectivity, and instrumental sensitivity in shaping SEP intensity profiles and the formation of IVA signatures.

\end{abstract}



\section{Introduction}

Research on Solar Energetic Particles (SEPs) has long focused on understanding how particles are accelerated to high energies during solar eruptions and subsequently transported through the heliosphere \citep[e.g.,][]{Desai2016, Reames2017, Reames2023}. Traditional models attribute large SEP events primarily to diffusive shock acceleration (DSA) at coronal mass ejection (CME)–driven shocks, whereby charged particles gain energy through repeated crossings of the shock front \citep[e.g.,][and references therein]{Vainio2018}. Decades of in situ observations have established the characteristic velocity‑dispersion signature of SEP onsets where higher‑energy particles arriving before lower‑energy ones and enabled techniques such as velocity‑dispersion analysis \citep[VDA; e.g.][]{Vainio2013, Laitinen2015} to infer particle release times and the path length traveled by SEPs from their source to the observers. Recent observations by Solar Orbiter \citep{Muller2020} and Parker Solar Probe \citep{Fox2016} close to the Sun, have uncovered an interesting new characteristic of some SEP events: an apparent inverse velocity dispersion \citep[IVD: e.g.][]{Sarris1976} or arrival of SEPs in which the first observed higher-energy particles arrive later than the lower ones in a way that concave (``nose-like'') shape appears in their energy spectrogram at the events' onset \citep[e.g.,][]{Cohen2024}. Energetic particle events with these dispersion characteristics has been first observed upstream of Earth's magnetosphere \citep[e.g.][]{Ipavich1981, Anagnostopoulos1986, Sarris1987}. \citet{Sarris1987} showed that the observed IVD for these events of magnetospheric origin, was most likely a propagation rather than a source effect \citep[e.g., fermi acceleration, see][]{Ipavich1981}. Following \cite{Xu2026arXiv}, in this study we use the term ``inverse velocity arrival'' (IVA) SEP events to differentiate from those events of magnetospheric origin that exhibit different spectral characteristics and probably physical mechanisms in their formation.

A prominent IVA-SEP event with a short duration of the IVA feature ($\sim$30 min) was observed close to the Sun by Parker Solar Probe (at $\sim$15 R$_{\odot}$) on 2022 September 5, with a nose energy\footnote{The energy that corresponds to the first-arriving particles.} for protons around 1~MeV \citep{Cohen2024}. For this event, \citet{Kouloumvakos2025} showed that the SEP release time was significantly delayed with respect to the onset of the parent eruption. The derived shock properties from 3D modeling suggested that the observed IVA could be attributed to an ongoing particle acceleration process at an initially weak region of the shock, which gradually strengthened. \cite{Xiaohang2025} further investigated this event and performed a simulation of particle acceleration and transport from the shock to Parker Solar Probe which suggested that the SEPs likely originated from a time-dependent DSA process at the flank of the expanding shock wave that intercepted Parker Solar Probe. The role of DSA in producing IVA signatures has also been explored theoretically by \cite{Yuncong2025} using Solar Orbiter observations.

Solar Orbiter has also observed many IVA-SEP events at distances between 0.3 and 1~au \citep{Allen2026, Yuncong2025}. For these events the IVA part has a significantly longer duration than that of the 2022 September 5 SEP event observed by Parker Solar Probe. \citet{Ding2025} from the analysis of one of these events observed on 2022 June 7 by Solar Orbiter suggested that the evolving magnetic connectivity between the spacecraft and the non-uniform propagating shock played a key role in the observation of a long-lasting IVA feature. In this case, the magnetic connection of the spacecraft to the evolving shock wave changed towards more efficient acceleration sites as the shock propagated farther from the Sun and thus producing the IVA effect.

Understanding when the IVA-SEPs arise in some events, whereas in some other do not, can help us to increase the knowledge about shock acceleration processes and thus improve our conceptual framework of the time-dependent DSA processes at fast evolving shocks. Both the evolving shock properties and magnetic connectivity between shock and spacecraft \citep{Kouloumvakos2019}, as well as the transport of SEPs to the spacecraft \citep[e.g.][]{Laitinen2015} determine the shape of the SEP intensity-time profiles \citep[also see][]{Cane1988}. \citet{Ding2025} was the first to suggest that the west-flank connectivity to the shock could play a central role for the formation of the IVA-SEP events. In addition, \citep{Allen2026} suggested that there is a spacial preference in observation of IVA-SEP events when the footpoints of the field lines connecting spacecraft to the Sun lie westward of the associated source (flare) location. Both studies suggest that an important factor is whether magnetic connection between the observers and the shock is established first on the western flank of the expanding shocks.

In this study, we examine the shock wave characteristics for IVA-SEP events. We provide a list of these events which were observed by Solar Orbiter or Parker Solar Probe, and modeled the associated shocks. Then we derive the shock parameters along the magnetic field lines connected to the observers. Our primary objective is to identify any systematic trends that could help explaining the formation of IVA signatures and thus establish a connection between the observed SEP properties and the corresponding shock characteristics. This approach aims to improve our understanding of where and when the observed IVA-SEPs originate along the expanding shock front and whether any relationships exist between the SEP properties and the shock dynamics.

\section{Observations and Modeling}

\subsection{SEP events exhibiting IVA}

For this study, we used SEP observations, below 1~au, from Solar Orbiter and Parker Solar Probe. Specifically, from Solar Orbiter we used SEP data from the Energetic Particle Detector suite \citep[EPD;][]{Rodriguez-Pacheco2020, Wimmer2021}, which includes the Supra-Thermal Electron and Proton sensor (STEP), the Electron Proton Telescope (EPT), the High Energy Telescope (HET) and the Suprathermal Ion Spectrograph (SIS). The combination of all these instruments, provides data for protons and heavier ions, from a few keV nucleon$^{-1}$ to 100 MeV nucleon$^{-1}$. From Parker Solar Probe we used SEP data from the Integrated Science Investigation of the Sun (IS$\odot$IS) instrument suite \citep{McComas2016} which includes the Energetic Particle Instruments, EPI-Lo and EPI-Hi \citep{Hill2017, Wiedenbeck2017} that measure ions at an energy range from $\sim$20~keV nucleon$^{-1}$ to $\sim$80~MeV nucleon$^{-1}$.

We surveyed SEP data from the two spacecraft, from the beginning of each mission (August 2018 and February 2020, for Parker Solar Probe and Solar Orbiter respectively) to March 2025, and we registered events, by visual inspection, that exhibit IVA. As we mentioned earlier, in the energy spectrograms this appears as a concave shape that is formed at the transition between the part where the SEPs follow a normal velocity dispersion (i.e. higher-energy particles arrive earlier than the lower-energy ones) and the part where the higher‑energy particles arrive later than lower‑energy ones above a transition energy (E$_t$), i.e. the energy where the IVA starts to form.

\begin{figure}[!ht]
    \centering
    \includegraphics[width=0.48\textwidth]{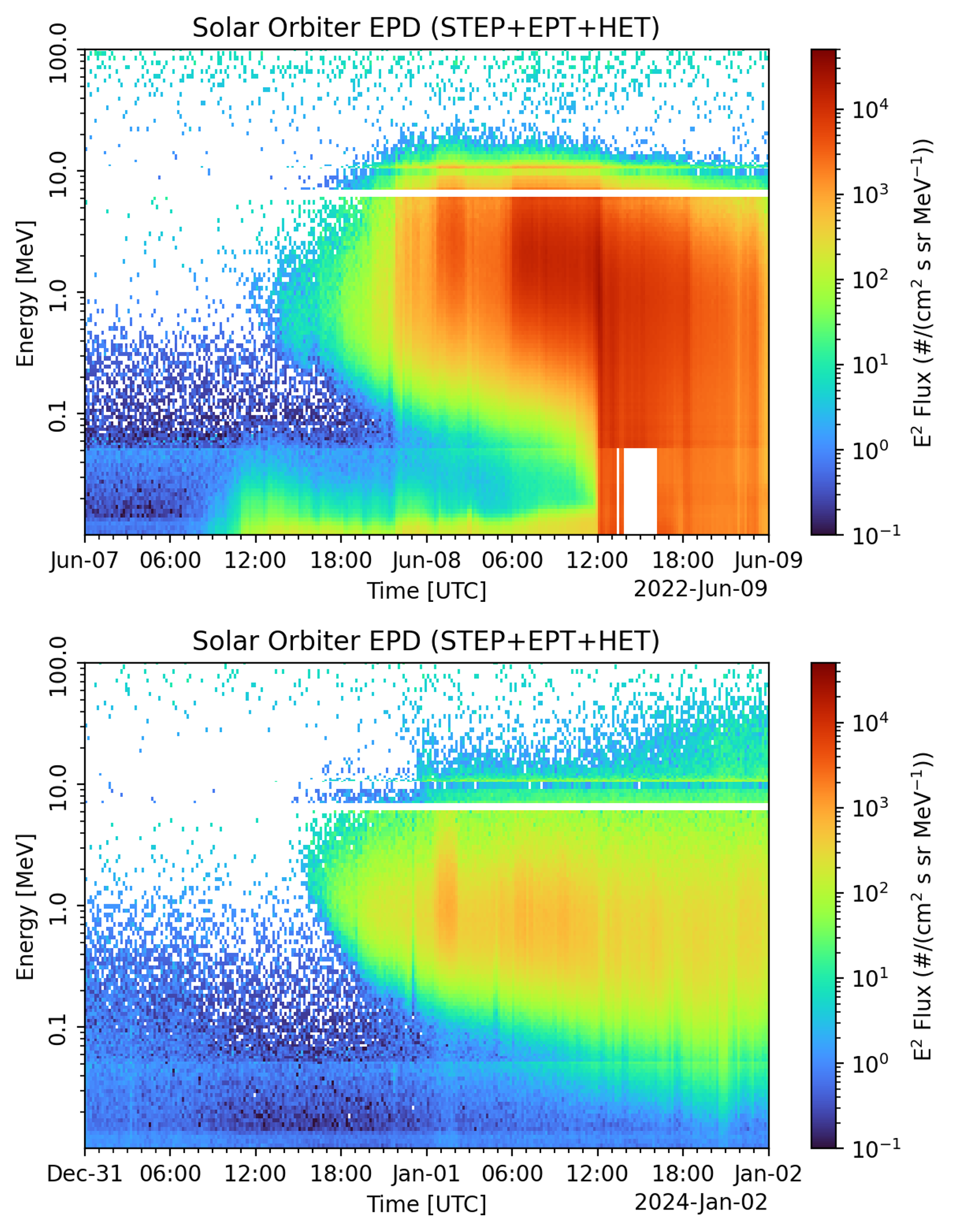}
    \caption{Observations of solar energetic protons for two IVA-SEP events observed by Solar Orbiter EPD. The top panel shows the energy spectrogram for the 2022 June 7 SEP event  and the bottom panel for the 2024 December 31 event. The two spectrograms cover an energy range from 10 keV to 100 MeV utilizing observations from EPD STEP, EPT, and HET instruments, and the fluxes have been averaged at a 10-minute cadence. For EPT and HET, only the sunward-looking telescopes were used. The color bar is multiplied by the square of the particle energies. }
    \label{fig:SEP_spectrograms}
\end{figure}

\begin{figure}[!ht]
    \centering
    \includegraphics[width=0.40\textwidth]{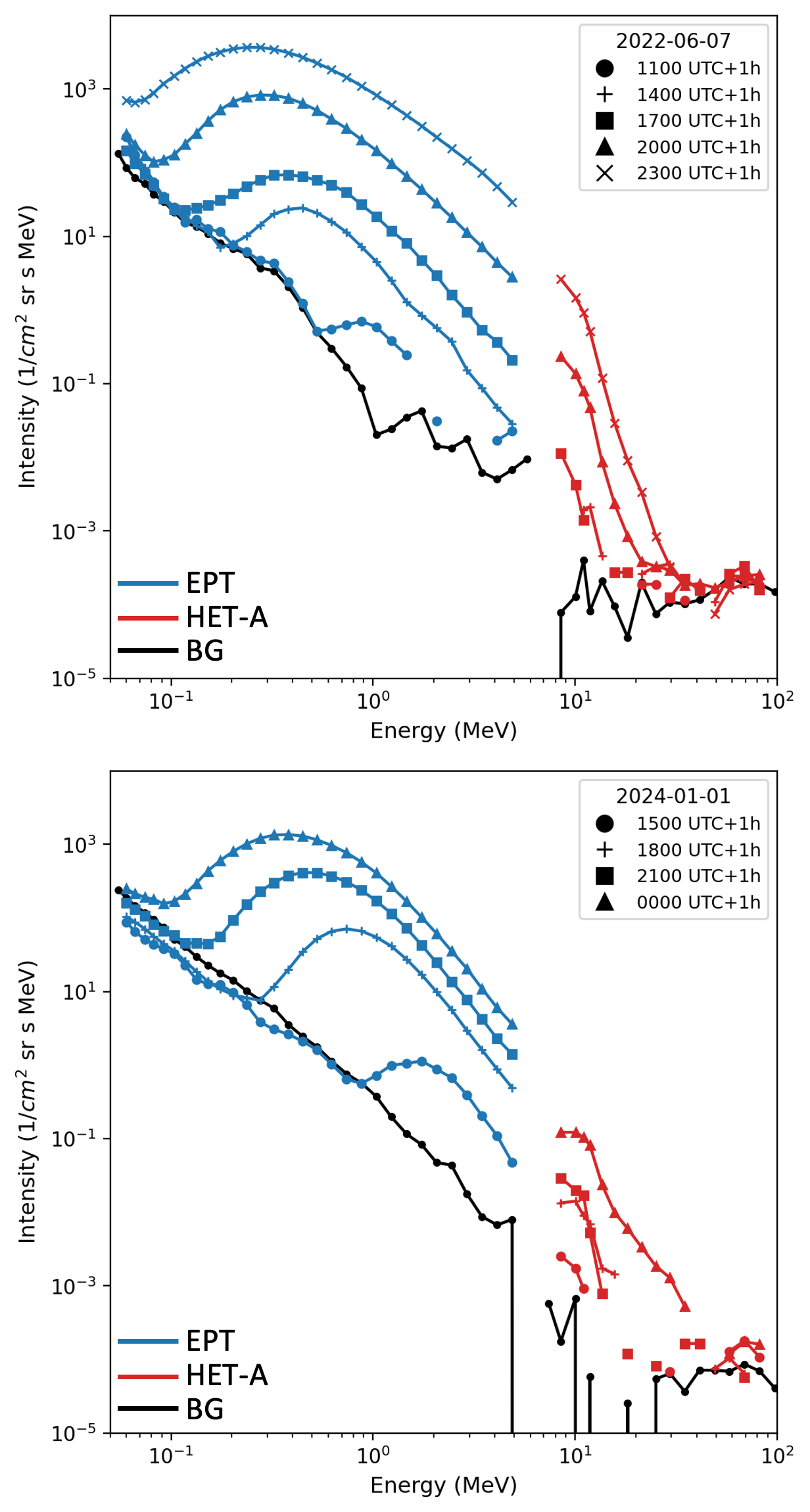}
    \caption{Temporal evolution of the SEP hourly-averaged energy spectrum for two IVA-SEP events observed by Solar Orbiter EPD. The top panel shows the energy spectra for the 2022 June 7 SEP event and the bottom panel for the 2024 December 31 event. The energy spectra for the top(bottom) panel, cover a range from $\sim$60~keV to $\sim$100~MeV during four(three) time intervals separated every three hours and using the sunward-looking telescopes from EPT (blue curves) and HET (red curves) instruments. In both panels, the SEP background determined a few hours before the event is presented with the black curves.}
    \label{fig:SEP_spectrum}
\end{figure}

Figure~\ref{fig:SEP_spectrograms} shows an example of two IVA-SEP events. These two events were observed by Solar Orbiter at similar heliocentric distances, on 2022 June 7 at 0.96~au (top panel) and 2024 December 31 at 0.95~au (bottom panel). The two events exhibit different characteristics. The 2022 June 7 SEP event has a E$_t$ at around 0.8~MeV and it is not until 12 hours later that the particles with the maximum energy ($\sim$20~MeV) reached the spacecraft. On the other hand, for the 2024 December 31 event, E$_t$ is $\sim$2 MeV, and the maximum-energy protons (at $\sim$60 MeV) are observed with a delay of almost one day with respect to those with energy E$_t$. For both events we see that the SEP intensities have a very gradual rise either below or above the E$_t$. Below the E$_t$, the normal velocity dispersion forms as expected. For any SEP event, instrument sensitivity and background affect the point at which the solar particle signal rises above any pre-existing background. Figure~\ref{fig:SEP_spectrum} illustrates the evolution of the energy spectra during the rise phase of the two events in Figure~\ref{fig:SEP_spectrograms}. The figure shows the pre-existing background spectrum and the energy spectra at different times as intensities rise reaching approximately one order of magnitude above the background. Because of this large signal, the time to maximum intensity is not significantly affected by the pre-existing background. While background and sensitivity effects can in principle influence the earliest onsets in weak events, we did not find evidence that such limitations are important for the intense events analyzed here. The spectral evolution in Figure~\ref{fig:SEP_spectrum} further shows that the spectrum is initially soft and hardens with time. This hardening indicates that the highest-energy particles reach their maximum intensity later than lower-energy particles, which may reflect continued evolution of the acceleration process.

In this survey we found 26 events, 21 for Solar Orbiter and five for Parker Solar Probe. We present the list of events in Appendix A (Table~\ref{tab:event_list}). The IVA-SEP events are observed at distances from 0.08 to $\sim$1~au, with almost one third of the events observed at heliocentric distances below 0.5~au (mean: 0.62~au). We do not find any clear trend of event occurrence with radial distance, which suggests that this does not play a role in their direction.

For each SEP event, we examined white-light coronagraph observations within approximately 12 hours before the SEP onset. In this interval, we identified candidate CMEs that could be responsible for the origin of the SEP events. In most cases, a single, clearly identifiable eruption (e.g., large-scale CME) was present near the SEP onset. In cases of multiple CMEs, we evaluated the timing and propagation direction of the CMEs relative to the nominal Parker spiral magnetic field footpoints of the observing spacecraft, to determine the most plausible association. Once the associated CME was identified, the associated flare was determined by linking the CME to its low-coronal source region using EUV imagery (SDO/AIA, STEREO-A/EUVI, Solar Orbiter/EUI). For the flare association we require temporal consistency between the CME lift-off, flare emission in EUV, and the flare time profiles using observations from the Spectrometer Telescope for Imaging X-rays \citep[STIX;][]{Krucker2020} and the X-ray Sensor (XRS) on board the GOES satellites.

For the IVA events studied here, we found that most of the associated CMEs have widths in plane-of-sky $>$120$^\circ$ and are characterized as halo (12/22) or partial-halo (8/22) events in the Coordinated Data Analysis Workshop (\href{https://cdaw.gsfc.nasa.gov/CME_list/}{CDAW}) SOHO/LASCO CME Catalog (see Table~\ref{tab:event_list}). Also from the flare associations, we found that for \textbf{15 events the associated flare was $<$M-class, for 7 events} was $\geq$M-class, and for four cases we could not determine the associated flare either because no data were available or the association was ambiguous due to multiple eruptions. Events without a clear flare and/or CME association were left out of the shock  (indicated by asterisks in Table~\ref{tab:event_list}). For the events \#2$^*$, 3$^*$, 7$^*$, multiple eruptions were present near the SEP events, so \textbf{we found it difficult} to conclude the solar eruption responsible for the IVAs. These three events and four other where the associated shock was too faint for 3D reconstruction (\#1$^*$, 5$^*$, 6$^*$) or the reconstruction was not possible due to data limitations (\#4$^*$) \textbf{were left out of the analysis}.

\subsection{Shock 3D Reconstruction and Modeling}

\begin{figure}
    \centering
    \includegraphics[width=0.48\textwidth]{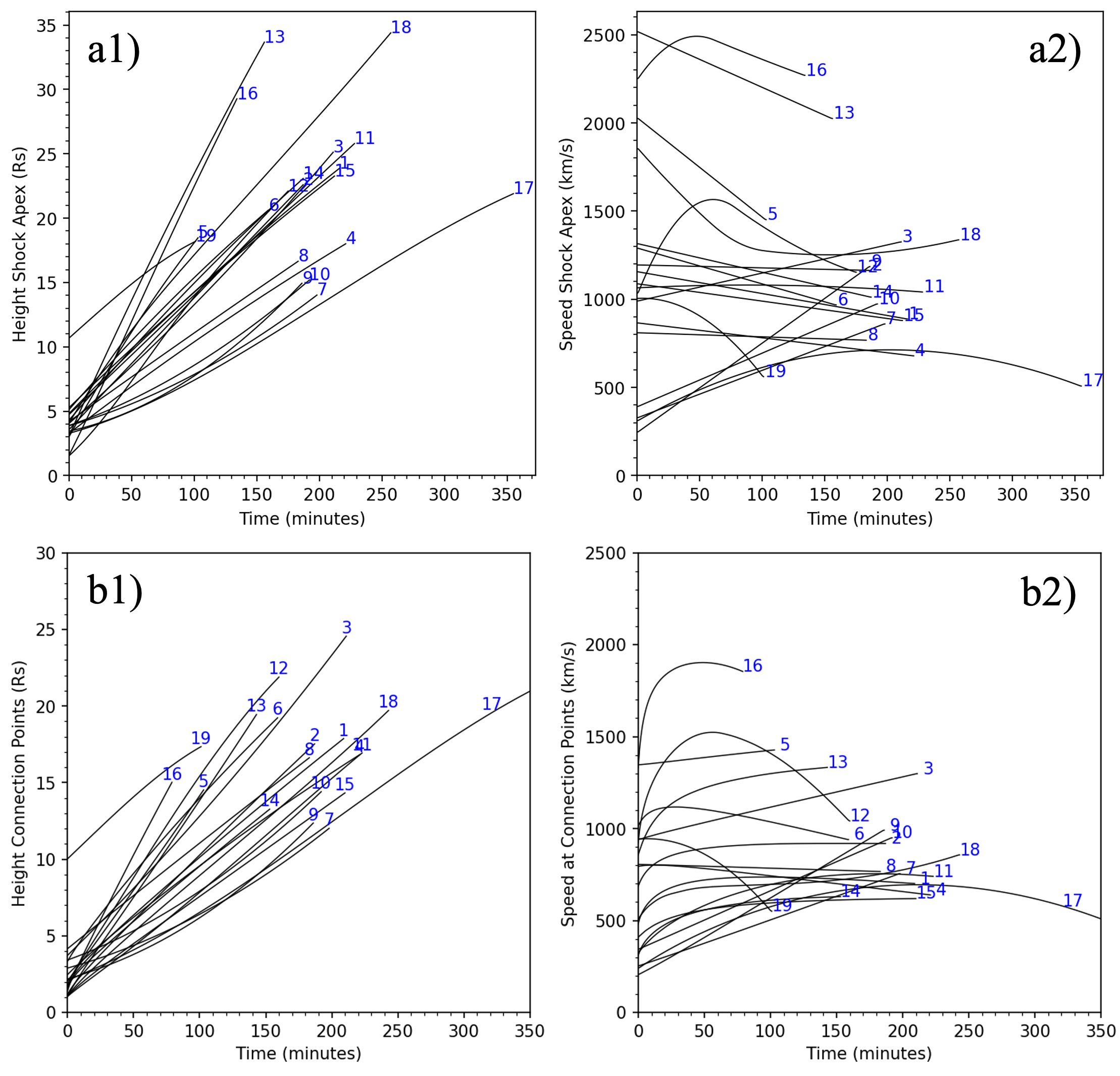}
    \caption{Kinematics of the 3D reconstructed shock waves associated with the IVA-SEP events. Panels a1) and b1) shows the height–time profile at the shock apex and the COBPOINTs, while panels a2) and b2) show the corresponding speed profiles. Time is measured relative to the start of the shock reconstruction. The labels shown are the IVA event \# in Table~\ref{tab:event_list}}.
    \label{fig:Shock_Kinematics}
\end{figure}

We performed a three‑dimensional (3D) reconstruction and kinematic modeling of the shock waves associated with the events using near‑simultaneous white‑light imagery provided by the coronagraphs of the Sun-Earth Connection Coronal and Heliospheric Investigation \cite[SECCHI;][]{Howard2008} and the Large Angle and Spectrometric Coronagraph \citep[LASCO;][]{Brueckner1995} from two vantage points where the Ahead spacecraft of the Solar Terrestrial Relations Observatory (STEREO-A) and the Solar and Heliospheric Observatory (SOHO) are located. Whenever possible we also used EUV data from STEREO‑A Extreme Ultraviolet Imager \citep[EUVI;][]{Wuelser2004} and SDO Atmospheric Imaging Assembly \citep[AIA;][]{Lemen2012} (for near‑Earth observations) to reconstruct the pressure/shock waves close to the solar surface. We used the PyThea software package \citep{Kouloumvakos2022_Pythea} and we fitted an ellipsoidal geometric model to the observed shock front. This geometrical model has been extensively used to model the global large-scale structure of propagating shocks without, however, including strongly non‑spherical or corrugated morphologies. We provide more details about the 3D reconstruction in Appendix~\ref{append:3DRec}.

\begin{figure*}[ht!]
    \centering
    \includegraphics[width=0.99\textwidth]{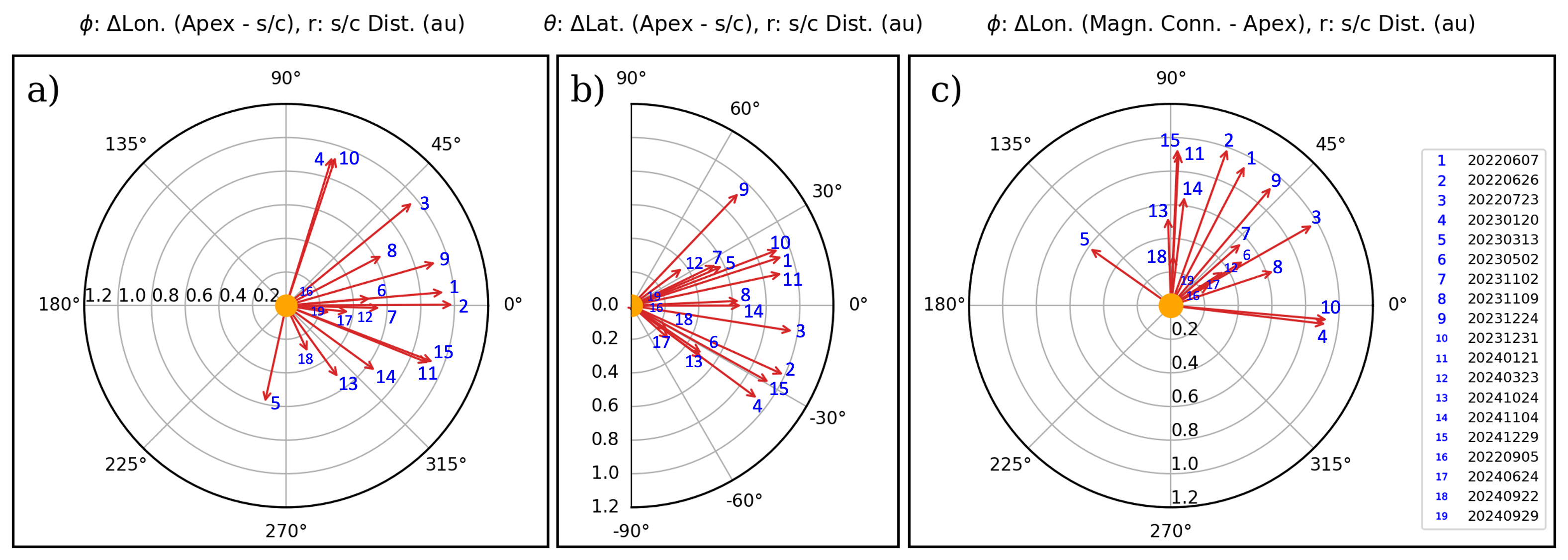}
    \caption{Panels (a) and (b) present polar plots showing the average longitudinal and latitudinal separation between the shock apex and the spacecraft. In panel (a), the angular values from 0$^\circ$ and increasing counterclockwise to 180$^\circ$ indicate that the shock apex propagates westward relative to the spacecraft. Conversely, angles measured clockwise from 0$^\circ$ indicate eastward propagation. In panel (b), angular values from 0$^\circ$ to 90$^\circ$ correspond to northward propagation, while those from 0$^\circ$ to -90$^\circ$ correspond to southward propagation. Panel (c) shows a polar plot similar to panel (a), but for the time average longitudinal separation between the shock apex and the spacecraft’s magnetic footpoints at the Sun, assuming a Parker spiral. The length of each arrow is scaled to the heliocentric distance of the spacecraft, at the time when the shocks are first reconstructed. The Sun is shown at the center of the plots, though not to scale.}
    \label{fig:Shock_Location_Polar}
\end{figure*}

From the 3D reconstructions we calculated the shock kinematics using a polynomial or spline fitting (depending on the quality of the 3D fittings) to derive the final height-time profiles of the shock apex and flanks with a uniform one minute cadence which we then used to perform the 3D model. Then we estimated the shock speed profiles from the time derivative of the height-time points. Using the results of the shock kinematics from the 3D reconstruction and magnetohydrodynamic (MHD) parameters of the background solar corona, we estimated the shock parameters in 3D during the shock evolution in the corona \citep[see][for further details of the shock modeling methods]{Rouillard2016, Kouloumvakos2019}. For the MHD parameters of the background corona, we used high-resolution MHD data from the Magnetohydrodynamic Algorithm outside a Sphere \citep[MAS;][]{Lionello2009} which provide estimates of plasma density and temperature in the corona \citep{Riley2011} from the solar surface to 30~R$_\sun$, which is the outer boundary of the coronal model. The MAS data are regularly provided by Predictive Science Inc.\footnote{\url{https://www.predsci.com/}} for each Carrington rotation. 

From the temporal evolution of the shock parameters in 3D, we calculated the evolution of the shock parameters at the magnetic field lines connected to the observers. In this study, we used a simple approach to calculate the full connectivity to the shock surface by taking nominal Parker spiral field lines connected to each observer and calculating the shock parameters at each COBPOINT (i.e., Connecting with the OBserver POINT following \cite{Heras1995}). When solar wind observations were available, we used the measured solar wind speed for the Parker spirals. The final shock parameters that we will present in the following figures, are mean values of the estimated parameter values at the COBPOINTs over the full temporal shock evolution that we modeled except if otherwise noted. In Table~\ref{tab:shock_param} we provide the shock parameters for each event, such as, the direction of the shock apex and the shock speed at the apex and the COBPOINTs.

\section{Results}

\subsection{Shock modeling results}

We started our analysis with the shock kinematics for each IVA-SEP events. Figures~\ref{fig:Shock_Kinematics}a1 and a2 show the time profiles of (a1) the height (heliocentric distance) of the shock apex and (a2) the shock speed at the shock apex for each event. Figures~\ref{fig:Shock_Kinematics}b1 and b2 show the time profiles of (b1) the height (or helioradii) of the COBPOINTs and (b2) the shock speed at the COBPOINT of each spacecraft for each event. From the figure we infer that most IVA-SEP events are associated with shocks with a mean shock speed of $\sim$1140~km/s at the apex at 5~R$_\sun$. Moreover, from the speed profiles of the shock apex, we found that most of the events (e.g., events \#6, \#11, \#15) undergo a deceleration, while others propagate at nearly constant speed or accelerate from $\sim$5~R$_\sun$ to $\sim$20~R$_\sun$ (i.e., events \#7, \#8, \#10).

Next we examined the shock kinematics at the COBPOINTs. Figure~\ref{fig:Shock_Kinematics}b1 shows that the first magnetic connections are established at heights below 5~R$_\sun$. Figure~\ref{fig:Shock_Kinematics}b2 shows that the shock speeds at the COBPOINT are lower than the apex speeds, which is expected since the CME-driven shock waves are mostly strongly driven near the apex \citep[e.g.,][]{Kwon2017, Jarry2023}. Moreover, the IVA-SEP events seem to be associated mainly with slower shocks at the COBPOINTs with a mean speed of $\sim$882~km/s. Lastly, we see that the shock speed at first connection is typically lower than the speed near the end of the modeling interval ($\sim$20-30~R$_\sun$ depending on the end time of the shock modeling). In most cases, the speed increases at the COBPOINTs for about 30 to 60~minutes, reaching a plateau afterwards (with some exceptions, e.g., events \#3, \#9, \#10, \#18). This is primarily due to changes in magnetic connection from the flanks of the shocks to the apex (which expands typically faster).

From the shock 3D reconstruction we determined the mean direction of propagation of the shock apex for each associated shock. Figures~\ref{fig:Shock_Location_Polar}a and b show the (a) longitudinal and (b) latitudinal separation angles between the apex shock direction and the spacecraft location in polar coordinates, where the length of the red arrows indicates the heliocentric radial distance of the spacecraft. Longitudinal angles between 0$^{\circ}$ and 90$^{\circ}$ (0$^{\circ}$ and 270$^{\circ}$) indicate that the shock apex propagates westward (eastward) relative to the spacecraft. From this analysis, we found that most events cluster within a $\pm$45$^\circ$ for both the longitudinal and latitudinal separation, suggesting that the shocks associated with IVA-SEP events propagate directly toward the spacecraft. 

Next, we considered the magnetic connectivity of the observers to the shock front by examining the location of the magnetic connections with respect to the apex of the shock. Figure~\ref{fig:Shock_Location_Polar}c is a polar plot of the longitudinal separation angle between the shock apex and the spacecraft's magnetic footpoints at the Sun computed assuming a nominal Parker spiral. From this analysis we found that most of the arrows are clustered between $\sim$40$^\circ$ to $\sim$80$^\circ$ which implies that the observers are primarily connected western of the shock apex which is consistent with \citep{Allen2026} results. However, from this longitudinal separation angle alone, it is difficult to tell how close the connections are to the shock flanks. This depends on the shock width, the global 3D geometry, and the full temporal evolution of the COBPOINTs location on the shock surface, so we explored this further. 

\begin{figure}
    \centering
    \includegraphics[width=0.45\textwidth]{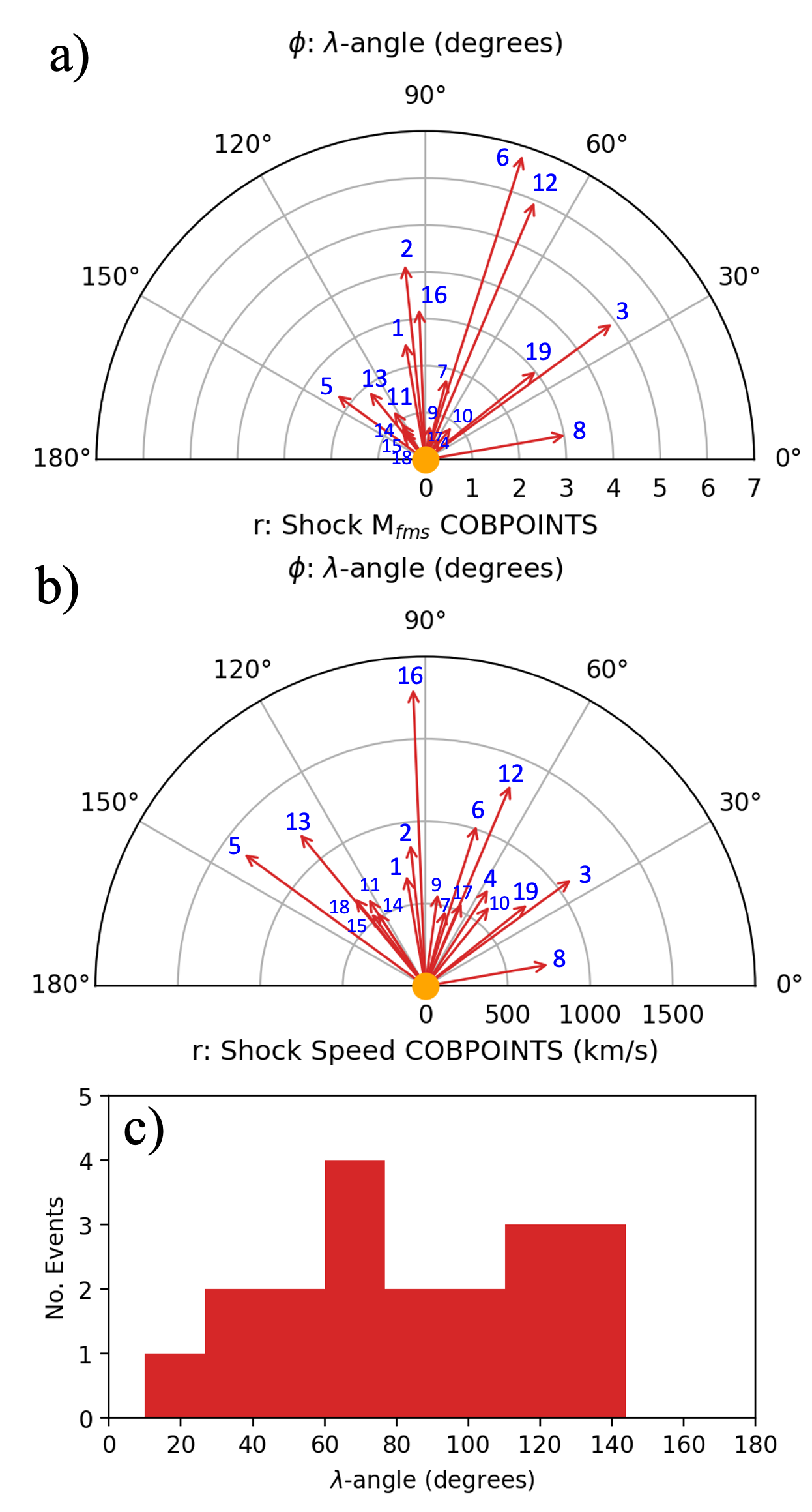}
    \caption{Panels a) and b) show polar plots of the mean central separation angle between the shock apex and the COBPOINTs. For panel a) the arrow lengths are scaled by the average fast magnetosonic Mach number, $M_{\text{fms}}$) and for panel b) by the shock speed, both measured at the COBPOINTs. Angular values near 0$^\circ$ indicate a magnetic connection close to the shock apex, values near $\sim$90$^\circ$ correspond to connections near the flanks, and values greater than 90$^\circ$ indicate connections below the flanks. Panel c) shows the distribution of the mean central separation angles. The labels shown are the IVA event \# in Table~\ref{tab:event_list} }
    \label{fig:Lambda_Polar}
\end{figure}

Figures~\ref{fig:Lambda_Polar}a and b are polar plots showing the $\lambda$-angle defined as the separation angle between the shock apex and the COBPOINT measured  with respect to the shock center (i.e., the center of the ellipsoid). This calculation involves both the longitude and latitude of the two points, so it is the angular distance of the COBPOINTs from the shock apex. \textbf{By definition, since the $\lambda$-angle is measured from the shock center, a value of 90$^\circ$ corresponds to a connection at the shock flanks.} Consequently, large values of $\lambda$ indicate magnetic connections to far-flanks ($\gg$90$^\circ$) or backside regions on the shock surface, \textbf{which can occur} in cases of globally extended shocks (e.g., for halo and partial-halo CMEs) where the $\lambda$-angle can be as large as 150$^\circ$--160$^\circ$.

\begin{figure*}
    \centering
    \includegraphics[width=0.98\textwidth]{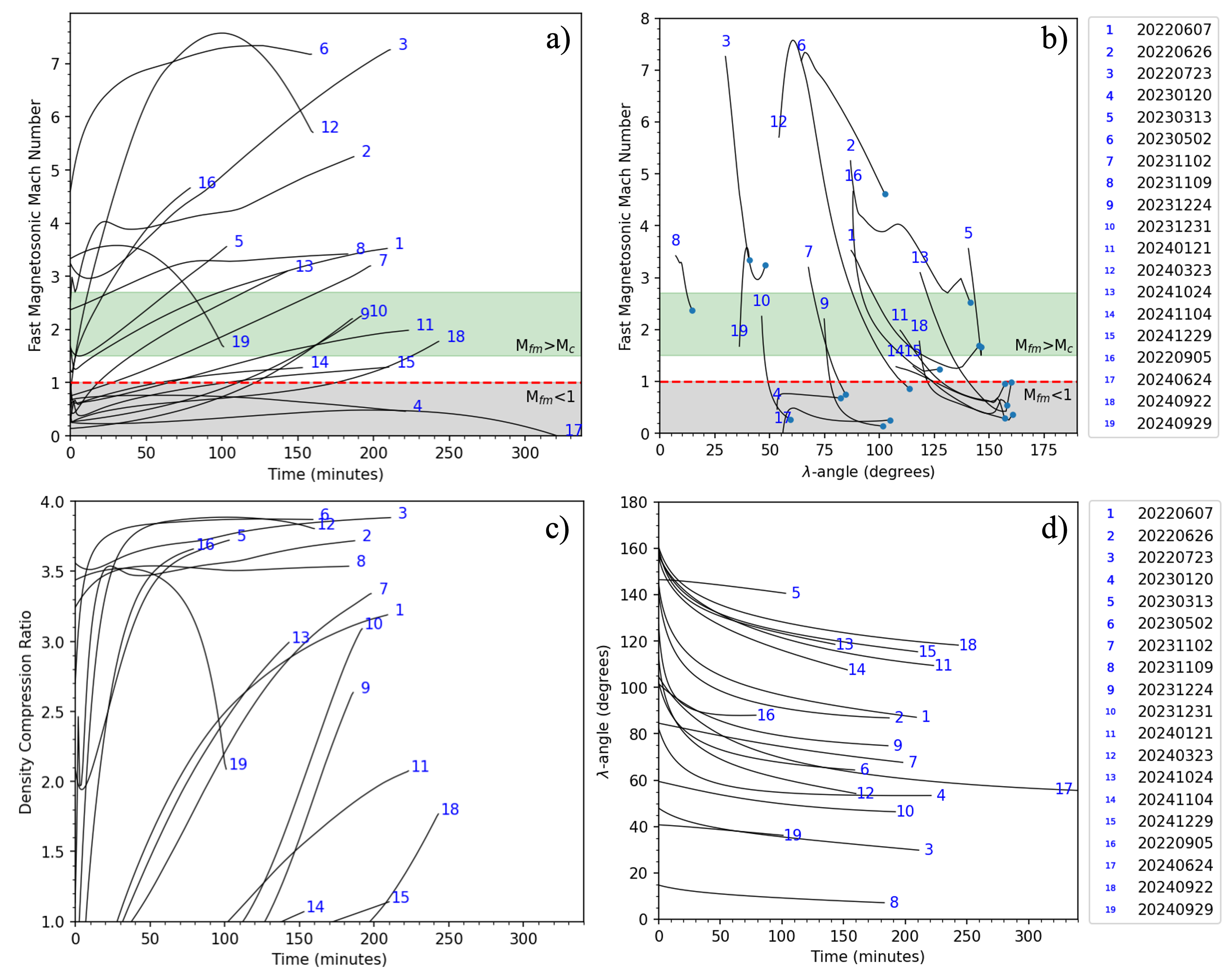}
    \caption{Evolution of the shock parameters along the magnetic field lines connected to the observers. Time is measured relative to the first modeled time when the reconstructed shock surface intersects the field lines connected to the observers. Panel a) shows the temporal evolution of the shock’s fast magnetosonic Mach number, $M_{\mathrm{fms}}$, and panel b) the evolution of $M_{\mathrm{fms}}$ as a function of the central separation angle, $\lambda$. Panel c) presents the temporal evolution of the density compression ratio at the COBPOINTs, and the lower-right panel shows the evolution of the $\lambda$-angle as a function of time. In the upper panels, the gray shaded area marks the range of $M_{\mathrm{fms}}$ values where no shock is formed, and the green shaded area the range where the shock is expected to become supercritical. In the panels showing the temporal evolution of the shock parameters, the start time corresponds to the beginning of the shock model, while in panel b) the same time is marked with blue symbols.}
    \label{fig:Mfm_profile}
\end{figure*}

\begin{figure}
    \centering
    \includegraphics[width=0.45\textwidth]{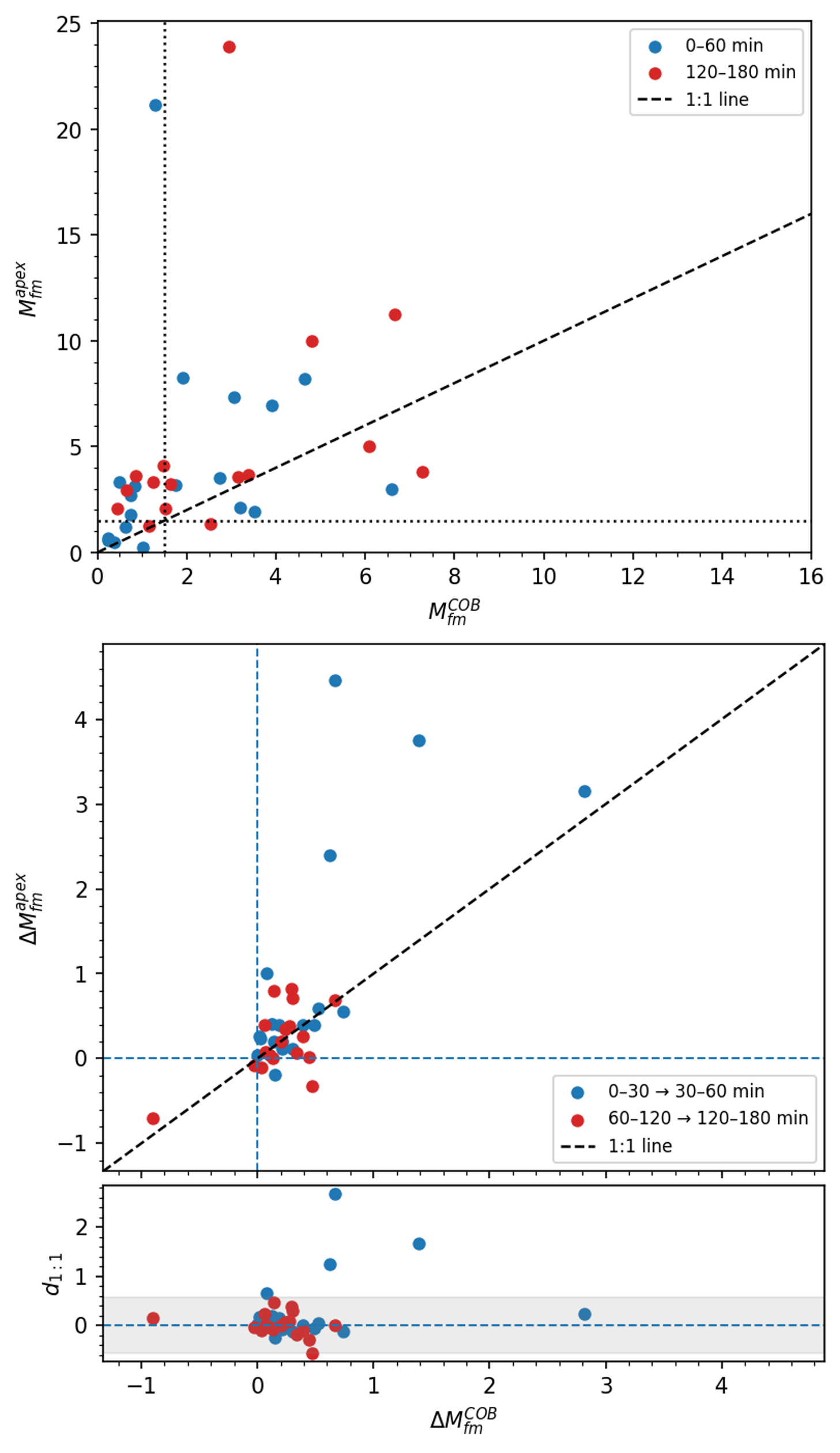}
    \caption{ Comparison of the $M_{\mathrm{fms}}$ evolution at the shock apex and at the COBPOINTs. Top panel: $M_{\text{fms}}^{\text{apex}}$ versus the $M_{\text{fms}}^{\text{COB}}$ for two representative time intervals after the start of the shock modeling (blue: 0--60 min; red: 120--180 min). The dotted horizontal and vertical lines mark the threshold separating weak/subcritical from stronger supercritical shocks. Bottom panel: Top axis show the changes in shock strength from $\Delta M_{\text{fms}}^{\text{apex}}$ versus $\Delta M_{\text{fms}}^{\text{COB}}$, between consecutive time windows (noted in the legend). Bottom axis show the deviation from proportional co-evolution, quantified as the signed distance from the 1:1 line (d$_{1:1}$) as a function of $\Delta M_{\text{fms}}^{\text{COB}}$. The gray shaded band highlights values close to the 1:1 relation. In both panels, the dashed black line shows the 1:1 relation. }
    \label{fig:Mfm_apex}
\end{figure}

For simplicity the vectors in Figure~\ref{fig:Lambda_Polar} show only the mean $\lambda$-angle during the shock evolution in our model domain (the full temporal evolution is shown later). The length of the arrows shows the mean values of the fast magnetosonic Mach number, $M_{\text{fms}}$ (panel a), or the shock speed (panel b) at the COBPOINTs over the full modeling interval. We found that most of the events are clustered between $\sim$30$^\circ$--110$^\circ$ (the shock flanks is at $\sim$90$^\circ$). Therefore, the magnetic connectivity of the spacecraft to the shock surface is primarily near the shock flanks for the IVA-SEP events analyzed here and suggests that the connection of the observers at the shock flanks plays an important role. Figure~\ref{fig:Lambda_Polar}c shows the distribution of the mean central separation angles and we have similar remarks as mentioned above. Considering also the mean shock parameters, $M_{\text{fms}}$ and speed as a function of $\lambda$-angle (e.g. Figures~\ref{fig:Lambda_Polar}a and b) we found no clear correlation. There is a trend for a few high-$M_{\text{fms}}$ shocks to appear for $\lambda$-angles above 90 degrees.


As a next step of this analysis we examined the full temporal evolution of the shock parameters. The top-row panels in Figure~\ref{fig:Mfm_profile} present the shocks’ $M_{\text{fms}}$ at the COBPOINT as a function of time and the $\lambda$-angle. The time profiles in panel a) reveal that the regions of the shocks initially magnetically connected to the observers are typically weak, with M$_{\text{fms}}\ll$1.5, i.e., sub-critical shocks. As the shocks evolve, these regions generally strengthen and become super-critical. In several cases, we observe that the shock formation is delayed (i.e., when $M_{\text{fms}}>1$), occurring approximately 0.5–2 hours after the first magnetic connection. This behavior where the initially weak shock regions become progressively super-critical characterizes the majority of the events analyzed here and it seems that this is another important property of the IVA-SEP events. Although notable exceptions exist, for instance, in two events (e.g., events \#4, \#17) the model shows no shock formation during the modeled time interval, while in three others (e.g., events \#3, \#6, \#19) the shocks are already super-critical at the moment of the first connection. For the later three cases, a closer investigation shows that the lack of EUV data to model the shocks from their initiation in the low corona, is probably a reason why these shocks appear to be immediately super-critical at the first connection. On the other hand, for the two events that remain in the non-shock region for the full modeling interval, a possibility is that these cases eventually became shocks well above the domain that we modeled. However, their downward trend in the shock strength makes this scenario rather unlikely, suggesting instead that the driving disturbances in these cases may have dissipated before reaching the shock-formation threshold. These two events (\#4, \#17) need further investigation and a more sophisticated modeling to conclude why they exhibit this behavior. Limitations of the methods used (i.e. the assumed magnetic connectivity, the uncertainty in the properties background medium) may also be a reason for this behavior.

\begin{figure*}
    \centering
    \includegraphics[width=0.92 \textwidth]{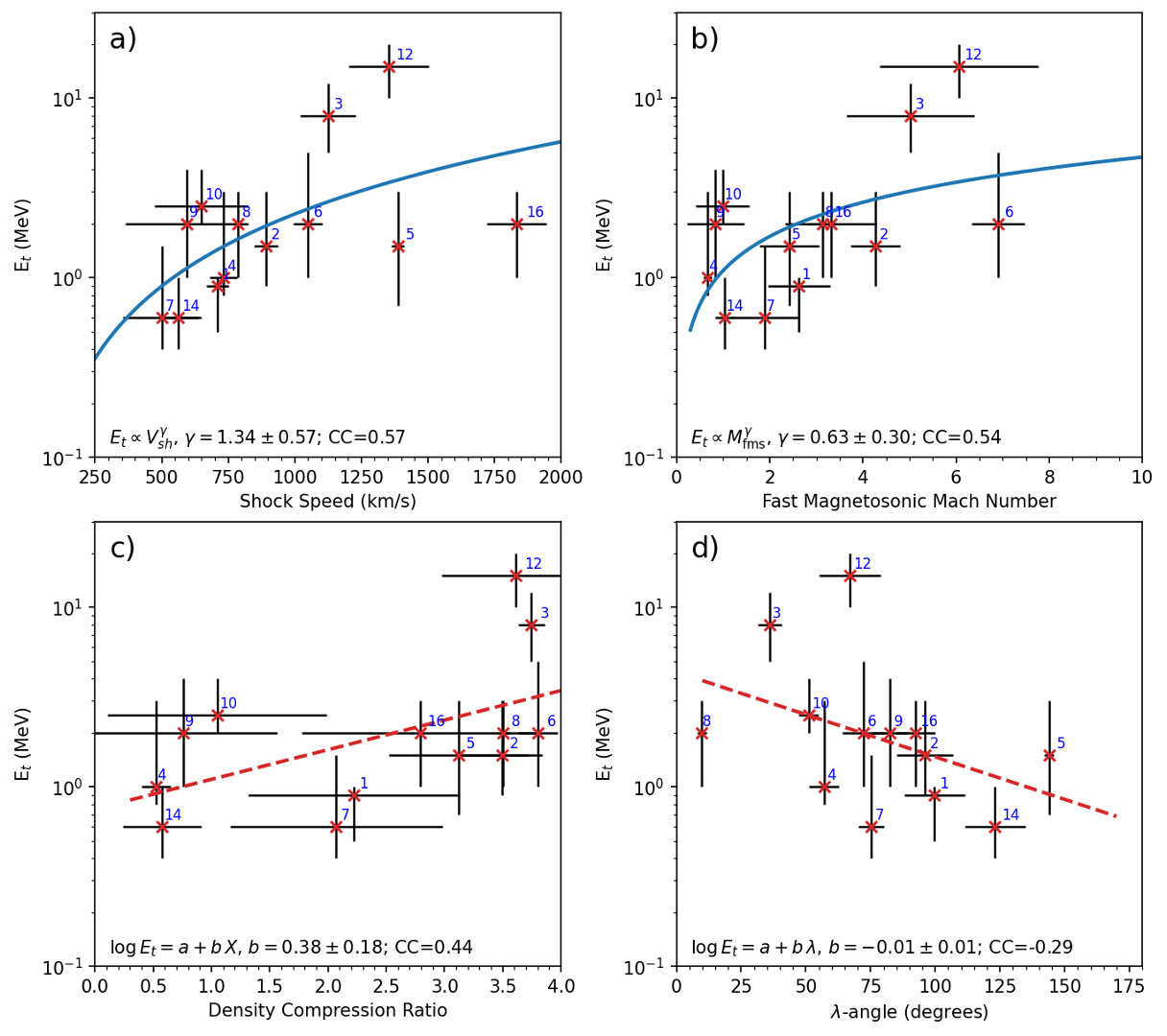}
    \caption{A connection between the transition energy, E$_t$, derived from the SEP observations and the mean shock parameters evaluated at the COBPOINTs. Panels show E$_t$ as a function of (a) shock speed (b) $M_{\text{fms}}$, (c) density compression ratio, and (d) $\lambda$-angle. The horizontal bars to the points show a proxy of the uncertainty of the visual determination of E$_t$ from the spectrograms and may span several energy channels in cases where the intensities are comparable. The vertical bars show the temporal standard deviation of the corresponding shock parameter over the modeled shock evolution. The best-fit relation shown in each panel (solid blue: power-law; dashed red: log-linear) and the corresponding Pearson correlation coefficient are reported in the lower-left corner. The labels shown are the IVA event \# in Table~\ref{tab:event_list}}.
    \label{fig:SEP_prop}
\end{figure*}

Figure~\ref{fig:Mfm_profile} b) shows the $M_{\text{fms}}$ as a function of the $\lambda$-angle. First, we see that the first magnetic connection to the observers (indicated by the blue symbols) is at regions near the flanks and far flanks of the shocks as noted earlier. Progressively the connectivity changes from the flanks towards the apex ($\lambda$-angle decreases), but the connections continue to be located closer to the shock flanks in most cases. Considering $M_{\text{fms}}$ as a function of $\lambda$-angle, we found that the first connected shock regions are initially weak shocks (or no shocks in some cases, e.g., in \#4 and \#17) and as the connectivity changes towards the apex, the shocks' strength increases and most shock regions become super-critical.

The bottom-row panels of Figure~\ref{fig:Mfm_profile} present the temporal evolution of the density compression ratio (panel c) and $\lambda$-angle (panel d) at the COBPOINTs. From panel c) we found that in most IVA events, the compression ratio increases with time, following a trend similar to that of $M_{\text{fms}}$. Initially, the shocks at the COBPOINTs exhibit relatively low density compression values (often close or below unity). The compression ratio in several cases, rises progressively and approaches values of $\sim$3--4, which indicates well-developed shocks. A few events, however, show persistently low compression ratios, consistent with what we showed for $M_{\text{fms}}$, where the initially connected regions correspond to weak shocks. For two events (\#4 and \#17), we find that no shock was formed ($X<1$). From panel d), we see that the $\lambda$-angle always decreases in time since the magnetic connection shifts from the flanks toward regions closer to the apex as the shock expands. For the majority of events, the change in connectivity is most rapid during the first hour of the shock evolution, after which the decrease in $\lambda$ becomes more gradual.

The overall evolution of the shock parameters at the COBPOINTs, suggests that this ``weak-to-strong'' (or ``subcritical-to-supercritical'') evolution of the shock is important for the formation of the IVAs and can be driven either by the global shock evolution (e.g., an intrinsic global temporal strengthening of the shock) or by changes in magnetic connectivity toward stronger regions of the expanding shock. To assess which of the two effects is more important, we compared the $M_{\text{fms}}$ at the shock apex ($M_{\text{fms}}^{\text{apex}}$) and at the COBPOINTs ($M_{\text{fms}}^{\text{COB}}$). In Figure~\ref{fig:Mfm_apex} (top panel) we compare the $M_{\text{fms}}^{\text{apex}}$ versus the $M_{\text{fms}}^{\text{COB}}$ for two different time windows during the shock evolution. We found that $M_{\text{fms}}^{\text{apex}} > M_{\text{fms}}^{\text{COB}}$ for the majority of events. This indicates that the magnetically connected regions are typically weaker than the shock apex at the same time which is consistent with the expectations from an initial magnetic connections to flank or backside regions. In addition, in several cases the shock at the apex becomes supercritical during its early evolution in the corona, while the corresponding regions at the COBPOINTs remain subcritical for a longer time. This shows that the transition to supercriticality generally occurs earlier at the apex than at the COBPOINTs. Importantly, the ordering $M_{\text{fms}}^{\text{apex}} > M_{\text{fms}}^{\text{COB}}$ persists when comparing the two time intervals which indicates that the relative hierarchy between apex and COBPOINTs is maintained as the shock evolves.

In Figure~\ref{fig:Mfm_apex} (bottom panel) we compare the changes in shock strength ($\Delta M_{\text{fms}}^{\text{apex}}$ and $\Delta M_{\text{fms}}^{\text{COB}}$) between consecutive time windows. The majority of events cluster close to the 1:1 line, indicating that when the shock apex strengthens, the shock regions at the COBPOINTs strengthen by a comparable amount. This near-proportional co-evolution implies that the temporal increase observed at the COBPOINTs largely follows the global strengthening of the shock. If connectivity changes were the dominant driver of the temporal evolution at the COBPOINTs, one would expect large variations in $\Delta M_{\text{fms}}^{\text{COB}}$ independent of $\Delta M_{\text{fms}}^{\text{apex}}$, or a systematic deviation from proportional co-evolution. Instead, the observed distribution shows that increases at the COBPOINTs closely track increases at the apex. A subset of events lies noticeably away from the 1:1 line so that the changes in shock strength at the apex and COBPOINTs are not strictly proportional. For these events, changes in the magnetic connectivity likely play a more important role in modulating the shock strength at the COBPOINTs. Such deviations are more pronounced during the early phase of the events, when connectivity changes are expected to occur more rapidly, as we showed earlier in Figure~\ref{fig:Mfm_profile} d), and become less significant at later times.
	
\subsection{Connection with SEP properties}

We also examined if there is a connection between the transition energy E$_t$ and the mean shock parameters at the COBPOINTs (Figure~\ref{fig:SEP_prop}). To characterize these relationships and avoid imposing an arbitrary functional form, we tested two simple parameterizations: (i) a power-law scaling, $\log E_t = \log A + \gamma \log P$ where P is either the shock speed or $M_{\rm fms}$, and (ii) a log-linear model, $\log E_t = a + b\,P$, which we adopt for the compression ratio and $\lambda$-angle. Overall, these choices allow us to emphasize the direction and relative strength of the trends while avoiding over-interpretation of the specific functional form in \textbf{this dataset of} limited size. For each parameter $P$ we quantified the strength of the association using the Pearson correlation coefficients (cc) and Spearman rank correlation ($\rho$).

Figure~\ref{fig:SEP_prop} a) shows $E_t$ as a function of the shock speed at the COBPOINTs. A positive trend is evident, with $E_t$ increasing with increasing $V_{\rm sh}$ with cc = 0.57 ($\rho$=0.53). For the $M_{\rm fms}$ at the COBPOINTs (Figure~\ref{fig:SEP_prop} b) we found a similar trend with cc =0.54 ($\rho$=0.46) indicating that the $E_t$ increase when the magnetically connected shock regions are stronger. For the density compression ratio (Figure~\ref{fig:SEP_prop} c) we found that a positive trend is present but the scatter is substantial and the correlation is weaker (cc =0.44, $\rho$=0.56) than the previous parameters. For the $\lambda$-angle we found only a very weak anti-correlation wit cc = --0.29 ($\rho$ = --0.60) indicating that $E_t$ may decrease with increasing angular distance from the shock apex. Table~\ref{tab:cc_summary} summarizes the Pearson and Spearman correlation coefficients (and associated p-values) between the transition energy $E_t$ and each of the mean shock parameters evaluated at the COBPOINTs. We note that the p-values for several cases are relatively high due to the low number of events and should be treated with caution.

\begin{table}[!h]
\centering
\caption{A summary of the Pearson correlation coefficients (cc) and Spearman rank correlations ($\rho$) for each parameter $P$ with respect to $E_t$.}
\label{tab:cc_summary}
\renewcommand{\arraystretch}{1.15}
\begin{tabular}{c c c}
\hline\hline
Param. $P$ & Pearson cc (p-val.) & Spearman $\rho$ (p-val.) \\
\hline
$V_{\rm sh}$ & 0.57 (0.040) & 0.53 (0.060) \\
$M_{\rm fms}$ & 0.54 (0.058) & 0.46 (0.113) \\
$X$ & 0.44 (0.133) & 0.56 (0.046) \\
$\lambda$ & --0.29 (0.332) & --0.60 (0.031) \\
\hline
\end{tabular}
\end{table}

\section{Discussion and Conclusion} \label{sec:Discussion}

In this study, we investigated IVA-SEP events observed, below 1~au, by Solar Orbiter and Parker~Solar~Probe. We compiled a list of 26 events, with 19 fully analyzed for their shock properties, where the IVA could be clearly identified in the SEP energy spectrograms at the onset of the events. Our analysis reveals that these events exhibit diverse characteristics in their transition energy. The energy spectra for a few cases we examined are typically soft and then progressively harden. This behavior is consistent with ongoing shock acceleration and/or a gradual shift in magnetic connectivity to stronger shock regions. Trying to differentiate both processes is difficult since the evolving propagating shocks imply that not only the shock conditions for particle acceleration vary, but also the observers do not establish magnetic connection with the same region of the shock front but sample different portions of the shock front as it propagates away from the Sun.

\cite{Allen2026} showed that SEP events with delayed high-energy proton onsets tend to occur when the magnetic connection of the observer lies westward of the associated flare. Here, we found a clear trend for the IVA-SEP events to predominantly occur when the spacecraft is magnetically connected western from apex of the associated shocks. This suggests that magnetic connectivity to the shock plays a predominant role in the formation of these SEP events confirming what outlined and predicted by \cite{Ding2025}. Previous studies have also shown that SEP events observed westward of the flare site tend to exhibit more gradual intensity profiles compared with those observed from eastern longitudes \citep[e.g.,][]{Cane1988, Rodriguez-Garcia2023}. This likely contributes to the observed spatial preference in the occurrence of IVA-events, as we explained earlier.

Extending this analysis to the analysis of the central separation angle seems to further confirm that overall a connection to the shock flanks favors the occurrence of these events. An additional contributing factor is the evolution of magnetic connectivity during the shock expansion. The analysis of the temporal evolution of the central separation angle seems to confirm that the overall evolving connection from the shock flanks to the apex possibly contributes to the occurrence of IVA events (e.g. see Appendix~\ref{append:model}).

Therefore, the predominance of IVA-SEP events associated with western magnetic connections likely reflects this evolution of both shock strength and magnetic connectivity changes. The gradual shift of field lines toward stronger shock regions naturally leads to a progressive increase in the maximum SEP energy which is a defining feature of the IVA-SEP events as shown in \cite{Allen2026}. However, this does not excludes the possibility that some IVA-SEPs may also form from regions with eastern connections, especially when a spacecraft observes close to the Sun. In this case, the changes in connectivity become less important and only the evolution the shock properties and the characteristics of time-dependent DSA can contribute to the formation of the IVA \citep[][]{Xiaohang2025}.

To further investigate the role of the shock properties in the formation of the IVA-SEP events, we performed detailed 3D modeling of the shocks. By analyzing the reconstructed shock geometry and evaluating the magnetic connectivity between the shock and the observers, we found that the majority of the IVA-SEP events are associated with magnetic connections to the flanks of the expanding shocks. This finding highlights the importance of the connections to the shock flanks in the formation of these spectral features as explained earlier. The analysis of the shocks' kinematics revealed that the regions magnetically connected to the observers typically have moderate-to-low propagation speeds (mean: $\sim$882~km/s) considering that large high-energy SEP events exhibit speeds at the COBPOINTs well above 1000~km/s as in the events analyzed by  \cite{Kouloumvakos2019}. Notably, associated EUV waves were identified in only a few events, suggesting that the shocks responsible for these SEP events likely form above the low corona (e.g. $>$5~R$_\sun$). This supports the interpretation that the observed SEP signatures arise from higher-altitude shock acceleration processes, particularly along the flanks.

Analysis of the evolution of shock parameters along magnetic field lines connected to the observers indicates that, in most cases, the initial connections lie at the flanks of the shock fronts. Additionally, the modeling shows that these connected regions are typically weak—subcritical—shocks characterized by low Mach numbers. As the shocks propagate through the middle corona, magnetic connectivity tends to shift progressively from the flanks toward the apex of the shock. This evolution is accompanied by a general increase in Mach number, with many shocks transitioning to super-critical conditions over time. Consequently, these properties imply slow acceleration rates, possibly through DSA, so that the longer the magnetic field lines remain attached to super-critical shock regions, the higher the energies of the particles being accelerated \citep[e.g.,][]{McComas2006}.

Our analysis also showed that a few events remain weak shocks and never become super-critical which is an inconsistency. One plausible explanation is that the MAS model overestimates the local fast magnetosonic speed in these cases, so that the modeled Mach numbers remain subcritical even though the actual shocks may eventually become super-critical. We also note that even a compression wave can accelerate particles under some certain conditions \citep[e.g.,][]{Wilson2025} which may partially account for these events. In any case, further investigation is needed for these events.

The combined effect of evolving magnetic connectivity and changes in the background coronal environment --both of which influence the shock strength at the field lines-- appears to play a central role in the formation of the IVA-SEP events. This dynamic interplay likely governs the gradual hardening of the spectrum and the delayed onset of high-energy particles observed in these events. We also examined whether the temporal strengthening observed at the COBPOINTs is driven primarily by the global shock evolution or by changes in magnetic connectivity, by directly comparing the shock strength at the shock apex and at the COBPOINTs. This analysis suggests that the global shock evolution seems to govern largely the shock strengthening at the COBPOINTs, with connectivity effects acting as an important but secondary effect rather being the principal driver for the time frame of the shock evolution that we examined in this study.

Next we explored connections of the shock parameters with SEP properties. The transition energy shows a tendency to scale with the shock speed (with Pearson cc$=$0.57) along the magnetically connected field lines, consistent with expectations from DSA theory since the acceleration rate at the shock can be expressed as a function of the shock speed, density compression ratio, and diffusion coefficient \citep{Jokipii1987}. However, the observed correlation with shock speed is marginally statistically significant (p-value$=$0.042) at the 0.05 level, since the data sample is low (13 points) and there also is a considerable scatter. We found similar results, but with lower cc's, for the $M_{\text{fms}}$ and the shock compression ratio. Several factors may contribute to this spread of the points. First, uncertainties in the shock reconstruction and magnetic connectivity can significantly affect the derived shock speeds at the COBPOINTs. Additionally, the ambiguity in determining E$_t$ for certain events may introduce an additional scatter. Another plausible explanation is that E$_t$ depends not only on shock speed but also on other physical parameters, such as the upstream parallel and perpendicular diffusion coefficient, which likely varies from event to event. These variations could obscure a tighter underlying relationship between shock properties and the observed transition energy. Future studies can elucidate further this aspect, for example if this correlation persists when adding more events and how the E$_t$ is theoretically expected to depend on the shock and particle transport properties and the location of the observers. In a recent study, \cite{Xiaohang2025} showed that E$_t$ occurs at the energy when the combined duration of SEP transport and acceleration reaches its minimum. 

\section{Summary}

The key findings of this study can be summarized as follows:

\begin{itemize}
    \item The shocks associated with the IVA-SEP events have typically moderate-to-low propagation speeds (mean: $\sim$882~km/s) compared to other large high-energy SEP events in \cite[e.g.,][]{Kouloumvakos2019}.
    
    \item IVA-SEP events occur predominantly when the observers are initially magnetically connected to the flanks of CME-driven shocks and in particular to the western flank.

     \item During the shock expansion, magnetic connections shift from the shock flanks toward the apex, where shocks are stronger.

    \item At the COBPOINTs the connected shock regions are initially sub-critical but often evolving to super-critical strengths which is consistent with the gradual hardening and delayed high-energy particle signatures observed in IVA-SEP events.

    \item The transition energy E$_t$ generally correlates with the mean shock speed at the COBPOINTs, though there is scatter reflecting variations in the shock parameters, diffusion conditions, and the uncertainties for both the modeled parameters and the measurements.

    \item The combined evolution of shock strength, magnetic connectivity, and coronal environment governs the formation of IVA-SEP events. 

\end{itemize}

Our results suggest that intrinsic shock-related processes are contributors to the formation of the IVA features, highlighting the need for future integrated analyses to disentangle their respective roles. Moreover, future investigations that combine multi-point SEP observations with detailed characterization of the full shock evolution will help clarify how variations in connectivity and shock strength shape the observed spectral behavior of the SEPs in IVA and non-IVA SEP events. Ultimately, these could point toward a unifying scenario where the interplay between evolving shock properties, magnetic field topology, magnetic connectivity, and particle transport conditions govern together the appearance and evolution of IVA-SEP events, offering new insights into the broader mechanisms of particle acceleration and release in the heliosphere.

\begin{acknowledgments}

We thank the anonymous Referee for their careful review and constructive comments, which significantly improved the clarity and quality of this manuscript. The authors would like to thank the IS$\odot$IS and the PSP mission teams. A.K. acknowledges financial support from the NASA contract NNN06AA01C (Solar Orbiter SIS, Parker Solar Probe EPI-Lo). SR acknowledges funding from Johns Hopkins University Applied Physics Laboratory independent R\&D fund. A.K., E.P., and A.V. acknowledge financial support from NASA’s LWS grant 80NSSC25K0130. A.V. is supported by NASA’s grants 80NSSC24K0555 and 80NSSC22K1028. P.R. was supported by NASA (80NSSC22K0893, 80NSSC20C0187, and 80NSSC20K1285), NSF’s PREEVENTS program (ICER-1854790), and NRL (N00173-24-C-0004). I.C.J. acknowledges support from the Research Council of Finland (X-Scale, grant No.~371569). L.R.-G. acknowledges support through the European Space
Agency (ESA) research fellowship programme. Solar Orbiter is a mission of international cooperation between ESA and NASA, operated by ESA. The Suprathermal Ion Spectrograph (SIS) is a European facility instrument funded by ESA under contract number SOL.ASTR.CON.00004. Solar Orbiter post-launch work at JHU/APL is supported by NASA contract NNN06AA01C, at the Southwest Research Institute by NASA 80GFSC25CA035. Parker Solar Probe was designed, built, and is now operated by the Johns Hopkins Applied Physics Laboratory (JHU/APL) as part of NASA's Living with a Star (LWS) program (contract NNN06AA01C). We thank the STEREO: SECCHI SOHO: LASCO; SDO/AIA teams and Predictive Science Inc. for providing the data used in this study. The STEREO SECCHI data are produced by a consortium of RAL (UK), NRL (USA), LMSAL (USA), GSFC (USA), MPS (Germany), CSL (Belgium), IOTA (France), and IAS (France). SOHO is a mission of international cooperation between ESA and NASA. The SDO/AIA data used are courtesy of SDO (NASA) and the AIA consortium. Some data processing for this research was carried out using version 7.1.0 of the SunPy open-source software package \citep{SunPy2020}. This research has made use of PyThea v1.1.0, an open-source and free Python package to reconstruct the 3D structure of CMEs and shock waves \citep{Kouloumvakos2022_Pythea}.

\end{acknowledgments}

\bibliography{sample7}{}

@ARTICLE{Allen2026,
	author = {{Allen}, R. C. and {Ho}, G. C. and {Mason}, G. M. and {Ding}, Z. and {Walker}, M. H. and {Kouloumvakos}, A. and {Wimmer-Schweingruber}, R. F. and {Rodriguez-Pacheco}, J. and {Vines}, S. K. and {Filwett}, R. J. and {Xu}, Z. and {Cohen}, C. M. S.},
	title = {Delayed maximum energy solar energetic particle events - Statistical analysis from Solar Orbiter},
	DOI= "10.1051/0004-6361/202557079",
	url= "https://doi.org/10.1051/0004-6361/202557079",
	journal = {\aap},
	year = 2026,
	volume = 705,
	pages = "A126",
}

@ARTICLE{Anagnostopoulos1986,
       author = {{Anagnostopoulos}, G.~C. and {Sarris}, E.~T. and {Krimigis}, S.~M.},
        title = "{Magnetospheric origin of energetic (at least 50 keV) ions upstream of the bow shock - The October 31, 1977, event}",
      journal = {\jgr},
         year = 1986,
        month = mar,
       volume = {91},
        pages = {3020-3028},
          doi = {10.1029/JA091iA03p03020},
       adsurl = {https://ui.adsabs.harvard.edu/abs/1986JGR....91.3020A}
}

@ARTICLE{Brueckner1995,
       author = {{Brueckner}, G.~E. and {Howard}, R.~A. and {Koomen}, M.~J. and {Korendyke}, C.~M. and {Michels}, D.~J. and {Moses}, J.~D. and {Socker}, D.~G. and {Dere}, K.~P. and {Lamy}, P.~L. and {Llebaria}, A. and {Bout}, M.~V. and {Schwenn}, R. and {Simnett}, G.~M. and {Bedford}, D.~K. and {Eyles}, C.~J.},
        title = "{The Large Angle Spectroscopic Coronagraph (LASCO)}",
      journal = {\solphys},
         year = 1995,
        month = dec,
       volume = {162},
       number = {1-2},
        pages = {357-402},
          doi = {10.1007/BF00733434},
       adsurl = {https://ui.adsabs.harvard.edu/abs/1995SoPh..162..357B}
}

@ARTICLE{Cane1988,
       author = {{Cane}, H.~V. and {Reames}, D.~V. and {von Rosenvinge}, T.~T.},
        title = "{The role of interplanetary shocks in the longitude distribution of solar energetic particles}",
      journal = {\jgr},
         year = 1988,
        month = sep,
       volume = {93},
       number = {A9},
        pages = {9555-9567},
          doi = {10.1029/JA093iA09p09555},
       adsurl = {https://ui.adsabs.harvard.edu/abs/1988JGR....93.9555C}
}

@ARTICLE{Cohen2024,
       author = {{Cohen}, C.~M.~S. and {Leske}, R.~A. and {Christian}, E.~R. and {Cummings}, A.~C. and {de Nolfo}, G.~A. and {Desai}, M.~I. and {Giacalone}, J. and {Hill}, M.~E. and {Labrador}, A.~W. and {McComas}, D.~J. and {McNutt}, R.~L. and {Mewaldt}, R.~A. and {Mitchell}, D.~G. and {Mitchell}, J.~G. and {Muro}, G.~D. and {Rankin}, J.~S. and {Schwadron}, N.~A. and {Sharma}, T. and {Shen}, M.~M. and {Szalay}, J.~R. and {Wiedenbeck}, M.~E. and {Xu}, Z.~G. and {Romeo}, O. and {Vourlidas}, A. and {Bale}, S.~D. and {Pulupa}, M. and {Kasper}, J.~C. and {Larson}, D.~E. and {Livi}, R. and {Whittlesey}, P.},
        title = "{Observations of the 2022 September 5 Solar Energetic Particle Event at 15 Solar Radii}",
      journal = {\apj},
         year = 2024,
        month = may,
       volume = {966},
       number = {2},
          eid = {148},
        pages = {148},
          doi = {10.3847/1538-4357/ad37f8},
       adsurl = {https://ui.adsabs.harvard.edu/abs/2024ApJ...966..148C}
}

@ARTICLE{Desai2016,
       author = {{Desai}, Mihir and {Giacalone}, Joe},
        title = "{Large gradual solar energetic particle events}",
      journal = {Living Reviews in Solar Physics},
         year = 2016,
        month = sep,
       volume = {13},
       number = {1},
          eid = {3},
        pages = {3}
}

@ARTICLE{Ding2025,
       author = {{Ding}, Zheyi and {Wimmer-Schweingruber}, Robert F. and {Kollhoff}, Alexander and {K{\"u}hl}, Patrick and {Yang}, Liu and {Berger}, Lars and {Kouloumvakos}, Athanasios and {Wijsen}, Nicolas and {Guo}, Jingnan and {Pacheco}, Daniel and {Li}, Yuncong and {Temmer}, Manuela and {Rodriguez-Pacheco}, Javier and {Allen}, Robert C. and {Ho}, George C. and {Mason}, Glenn M. and {Xu}, Zigong and {Gunaseelan}, Sindhuja},
        title = "{Investigation of the inverse velocity dispersion in a solar energetic particle event observed by Solar Orbiter}",
      journal = {\aap},
         year = 2025,
        month = apr,
       volume = {696},
          eid = {A199},
        pages = {A199},
          doi = {10.1051/0004-6361/202553806},
archivePrefix = {arXiv},
       eprint = {2503.12522},
 primaryClass = {astro-ph.SR},
       adsurl = {https://ui.adsabs.harvard.edu/abs/2025A&A...696A.199D}
}

@ARTICLE{Fox2016,
       author = {{Fox}, N.~J. and {Velli}, M.~C. and {Bale}, S.~D. and {Decker}, R. and {Driesman}, A. and {Howard}, R.~A. and {Kasper}, J.~C. and {Kinnison}, J. and {Kusterer}, M. and {Lario}, D. and {Lockwood}, M.~K. and {McComas}, D.~J. and {Raouafi}, N.~E. and {Szabo}, A.},
        title = "{The Solar Probe Plus Mission: Humanity's First Visit to Our Star}",
      journal = {\ssr},
         year = 2016,
        month = dec,
       volume = {204},
       number = {1-4},
        pages = {7-48},
          doi = {10.1007/s11214-015-0211-6},
       adsurl = {https://ui.adsabs.harvard.edu/abs/2016SSRv..204....7F}
}

@ARTICLE{Heras1995,
       author = {{Heras}, A.~M. and {Sanahuja}, B. and {Lario}, D. and {Smith}, Z.~K. and {Detman}, T. and {Dryer}, M.},
        title = "{Three Low-Energy Particle Events: Modeling the Influence of the Parent Interplanetary Shock}",
      journal = {\apj},
         year = 1995,
        month = may,
       volume = {445},
        pages = {497},
          doi = {10.1086/175714},
       adsurl = {https://ui.adsabs.harvard.edu/abs/1995ApJ...445..497H}
}

@ARTICLE{Hill2017,
       author = {{Hill}, M.~E. and {Mitchell}, D.~G. and {Andrews}, G.~B. and {Cooper}, S.~A. and {Gurnee}, R.~S. and {Hayes}, J.~R. and {Layman}, R.~S. and {McNutt}, R.~L. and {Nelson}, K.~S. and {Parker}, C.~W. and {Schlemm}, C.~E. and {Stokes}, M.~R. and {Begley}, S.~M. and {Boyle}, M.~P. and {Burgum}, J.~M. and {Do}, D.~H. and {Dupont}, A.~R. and {Gold}, R.~E. and {Haggerty}, D.~K. and {Hoffer}, E.~M. and {Hutcheson}, J.~C. and {Jaskulek}, S.~E. and {Krimigis}, S.~M. and {Liang}, S.~X. and {London}, S.~M. and {Noble}, M.~W. and {Roelof}, E.~C. and {Seifert}, H. and {Strohbehn}, K. and {Vandegriff}, J.~D. and {Westlake}, J.~H.},
        title = "{The Mushroom: A half-sky energetic ion and electron detector}",
      journal = {Journal of Geophysical Research (Space Physics)},
         year = 2017,
        month = feb,
       volume = {122},
       number = {2},
        pages = {1513-1530},
          doi = {10.1002/2016JA022614},
       adsurl = {https://ui.adsabs.harvard.edu/abs/2017JGRA..122.1513H}
}

@ARTICLE{Howard2008,
       author = {{Howard}, R.~A. and {Moses}, J.~D. and {Vourlidas}, A. and {Newmark}, J.~S. and {Socker}, D.~G. and {Plunkett}, S.~P. and {Korendyke}, C.~M. and {Cook}, J.~W. and {Hurley}, A. and {Davila}, J.~M. and {Thompson}, W.~T. and {St Cyr}, O.~C. and {Mentzell}, E. and {Mehalick}, K. and {Lemen}, J.~R. and {Wuelser}, J.~P. and {Duncan}, D.~W. and {Tarbell}, T.~D. and {Wolfson}, C.~J. and {Moore}, A. and {Harrison}, R.~A. and {Waltham}, N.~R. and {Lang}, J. and {Davis}, C.~J. and {Eyles}, C.~J. and {Mapson-Menard}, H. and {Simnett}, G.~M. and {Halain}, J.~P. and {Defise}, J.~M. and {Mazy}, E. and {Rochus}, P. and {Mercier}, R. and {Ravet}, M.~F. and {Delmotte}, F. and {Auchere}, F. and {Delaboudiniere}, J.~P. and {Bothmer}, V. and {Deutsch}, W. and {Wang}, D. and {Rich}, N. and {Cooper}, S. and {Stephens}, V. and {Maahs}, G. and {Baugh}, R. and {McMullin}, D. and {Carter}, T.},
        title = "{Sun Earth Connection Coronal and Heliospheric Investigation (SECCHI)}",
      journal = {\ssr},
         year = 2008,
        month = apr,
       volume = {136},
       number = {1-4},
        pages = {67-115},
          doi = {10.1007/s11214-008-9341-4},
       adsurl = {https://ui.adsabs.harvard.edu/abs/2008SSRv..136...67H}
}

@ARTICLE{Ipavich1981,
       author = {{Ipavich}, F.~M. and {Galvin}, A.~B. and {Gloeckler}, G. and {Scholer}, M. and {Hovestadt}, D.},
        title = "{A statistical survey of ions observed upstream of the earth's bow shock: Energy spectra, composition, and spatial variations}",
      journal = {\jgr},
         year = 1981,
        month = jun,
       volume = {86},
       number = {A6},
        pages = {4337-4342},
          doi = {10.1029/JA086iA06p04337},
       adsurl = {https://ui.adsabs.harvard.edu/abs/1981JGR....86.4337I}
}

@ARTICLE{Jarry2023,
       author = {{Jarry}, Manon and {Rouillard}, Alexis P. and {Plotnikov}, Illya and {Kouloumvakos}, Athanasios and {Warmuth}, Alexander},
        title = "{Parametric study of the kinematic evolution of coronal mass ejection shock waves and their relation to flaring activity}",
      journal = {\aap},
         year = 2023,
        month = apr,
       volume = {672},
          eid = {A127},
        pages = {A127},
          doi = {10.1051/0004-6361/202245480},
archivePrefix = {arXiv},
       eprint = {2303.08663},
 primaryClass = {astro-ph.SR},
       adsurl = {https://ui.adsabs.harvard.edu/abs/2023A&A...672A.127J}
}

@ARTICLE{Jokipii1987,
       author = {{Jokipii}, J.~R.},
        title = "{Rate of Energy Gain and Maximum Energy in Diffusive Shock Acceleration}",
      journal = {\apj},
         year = 1987,
        month = feb,
       volume = {313},
        pages = {842},
          doi = {10.1086/165022},
       adsurl = {https://ui.adsabs.harvard.edu/abs/1987ApJ...313..842J}
}

@ARTICLE{Krucker2020,
       author = {{Krucker}, S{\"a}m and {Hurford}, G.~J. and {Grimm}, O. and {K{\"o}gl}, S. and {Gr{\"o}belbauer}, H. -P. and {Etesi}, L. and {Casadei}, D. and {Csillaghy}, A. and {Benz}, A.~O. and {Arnold}, N.~G. and {Molendini}, F. and {Orleanski}, P. and {Schori}, D. and {Xiao}, H. and {Kuhar}, M. and {Hochmuth}, N. and {Felix}, S. and {Schramka}, F. and {Marcin}, S. and {Kobler}, S. and {Iseli}, L. and {Dreier}, M. and {Wiehl}, H.~J. and {Kleint}, L. and {Battaglia}, M. and {Lastufka}, E. and {Sathiapal}, H. and {Lapadula}, K. and {Bednarzik}, M. and {Birrer}, G. and {Stutz}, St. and {Wild}, Ch. and {Marone}, F. and {Skup}, K.~R. and {Cichocki}, A. and {Ber}, K. and {Rutkowski}, K. and {Bujwan}, W. and {Juchnikowski}, G. and {Winkler}, M. and {Darmetko}, M. and {Michalska}, M. and {Seweryn}, K. and {Bia{\l}ek}, A. and {Osica}, P. and {Sylwester}, J. and {Kowalinski}, M. and {{\'S}cis{\l}owski}, D. and {Siarkowski}, M. and {St{\k{e}}{\'s}licki}, M. and {Mrozek}, T. and {Podg{\'o}rski}, P. and {Meuris}, A. and {Limousin}, O. and {Gevin}, O. and {Le Mer}, I. and {Brun}, S. and {Strugarek}, A. and {Vilmer}, N. and {Musset}, S. and {Maksimovi{\'c}}, M. and {F{\'a}rn{\'\i}k}, F. and {Koz{\'a}{\v{c}}ek}, Z. and {Ka{\v{s}}parov{\'a}}, J. and {Mann}, G. and {{\"O}nel}, H. and {Warmuth}, A. and {Rendtel}, J. and {Anderson}, J. and {Bauer}, S. and {Dionies}, F. and {Paschke}, J. and {Pl{\"u}schke}, D. and {Woche}, M. and {Schuller}, F. and {Veronig}, A.~M. and {Dickson}, E.~C.~M. and {Gallagher}, P.~T. and {Maloney}, S.~A. and {Bloomfield}, D.~S. and {Piana}, M. and {Massone}, A.~M. and {Benvenuto}, F. and {Massa}, P. and {Schwartz}, R.~A. and {Dennis}, B.~R. and {van Beek}, H.~F. and {Rodr{\'\i}guez-Pacheco}, J. and {Lin}, R.~P.},
        title = "{The Spectrometer/Telescope for Imaging X-rays (STIX)}",
      journal = {\aap},
         year = 2020,
        month = oct,
       volume = {642},
          eid = {A15},
        pages = {A15},
          doi = {10.1051/0004-6361/201937362},
       adsurl = {https://ui.adsabs.harvard.edu/abs/2020A&A...642A..15K}
}

@ARTICLE{Kouloumvakos2019,
       author = {{Kouloumvakos}, Athanasios and {Rouillard}, Alexis P. and {Wu}, Yihong and
         {Vainio}, Rami and {Vourlidas}, Angelos and {Plotnikov}, Illya and
         {Afanasiev}, Alexandr and {{\"O}nel}, Hakan},
        title = "{Connecting the Properties of Coronal Shock Waves with Those of Solar Energetic Particles}",
      journal = {\apj},
         year = "2019",
        month = "May",
       volume = {876},
       number = {1},
          eid = {80},
        pages = {80},
          doi = {10.3847/1538-4357/ab15d7},
       adsurl = {https://ui.adsabs.harvard.edu/abs/2019ApJ...876...80K}
}

@ARTICLE{Kouloumvakos2022_Pythea,
       author = {{Kouloumvakos}, Athanasios and {Rodr{\'\i}guez-Garc{\'\i}a}, Laura and {Gieseler}, Jan and {Price}, Daniel J. and {Vourlidas}, Angelos and {Vainio}, Rami},
        title = "{PyThea: An open-source software package to perform 3D reconstruction of coronal mass ejections and shock waves}",
      journal = {Frontiers in Astronomy and Space Sciences},
         year = 2022,
        month = sep,
       volume = {9},
          eid = {974137},
        pages = {974137},
          doi = {10.3389/fspas.2022.974137},
       adsurl = {https://ui.adsabs.harvard.edu/abs/2022FrASS...9.4137K}
}

@ARTICLE{Kouloumvakos2025,
       author = {{Kouloumvakos}, A. and {Wijsen}, N. and {Jebaraj}, I.~C. and {Afanasiev}, A. and {Lario}, D. and {Cohen}, C.~M.~S. and {Riley}, P. and {Mitchell}, D.~G. and {Ding}, Z. and {Vourlidas}, A. and {Giacalone}, J. and {Chen}, X. and {Hill}, M.~E.},
        title = "{Shock and SEP Modeling Study for the 2022 September 5 SEP Event}",
      journal = {\apj},
         year = 2025,
        month = feb,
       volume = {979},
       number = {2},
          eid = {100},
        pages = {100},
          doi = {10.3847/1538-4357/ada0be},
archivePrefix = {arXiv},
       eprint = {2501.03066},
 primaryClass = {astro-ph.SR},
       adsurl = {https://ui.adsabs.harvard.edu/abs/2025ApJ...979..100K}
}

@ARTICLE{Kwon2014,
       author = {{Kwon}, Ryun-Young and {Zhang}, Jie and {Olmedo}, Oscar},
        title = "{New Insights into the Physical Nature of Coronal Mass Ejections and Associated Shock Waves within the Framework of the Three-dimensional Structure}",
      journal = {\apj},
         year = 2014,
        month = oct,
       volume = {794},
       number = {2},
          eid = {148},
        pages = {148},
          doi = {10.1088/0004-637X/794/2/148},
       adsurl = {https://ui.adsabs.harvard.edu/abs/2014ApJ...794..148K}
}

@ARTICLE{Kwon2017,
       author = {{Kwon}, Ryun-Young and {Vourlidas}, Angelos},
        title = "{Investigating the Wave Nature of the Outer Envelope of Halo Coronal Mass Ejections}",
      journal = {\apj},
         year = 2017,
        month = feb,
       volume = {836},
       number = {2},
          eid = {246},
        pages = {246},
          doi = {10.3847/1538-4357/aa5b92},
       adsurl = {https://ui.adsabs.harvard.edu/abs/2017ApJ...836..246K}
}

@ARTICLE{Lemen2012,
       author = {{Lemen}, James R. and {Title}, Alan M. and {Akin}, David J. and {Boerner}, Paul F. and {Chou}, Catherine and {Drake}, Jerry F. and {Duncan}, Dexter W. and {Edwards}, Christopher G. and {Friedlaender}, Frank M. and {Heyman}, Gary F. and {Hurlburt}, Neal E. and {Katz}, Noah L. and {Kushner}, Gary D. and {Levay}, Michael and {Lindgren}, Russell W. and {Mathur}, Dnyanesh P. and {McFeaters}, Edward L. and {Mitchell}, Sarah and {Rehse}, Roger A. and {Schrijver}, Carolus J. and {Springer}, Larry A. and {Stern}, Robert A. and {Tarbell}, Theodore D. and {Wuelser}, Jean-Pierre and {Wolfson}, C. Jacob and {Yanari}, Carl and {Bookbinder}, Jay A. and {Cheimets}, Peter N. and {Caldwell}, David and {Deluca}, Edward E. and {Gates}, Richard and {Golub}, Leon and {Park}, Sang and {Podgorski}, William A. and {Bush}, Rock I. and {Scherrer}, Philip H. and {Gummin}, Mark A. and {Smith}, Peter and {Auker}, Gary and {Jerram}, Paul and {Pool}, Peter and {Soufli}, Regina and {Windt}, David L. and {Beardsley}, Sarah and {Clapp}, Matthew and {Lang}, James and {Waltham}, Nicholas},
        title = "{The Atmospheric Imaging Assembly (AIA) on the Solar Dynamics Observatory (SDO)}",
      journal = {\solphys},
         year = 2012,
        month = jan,
       volume = {275},
       number = {1-2},
        pages = {17-40},
          doi = {10.1007/s11207-011-9776-8},
       adsurl = {https://ui.adsabs.harvard.edu/abs/2012SoPh..275...17L}
}

@ARTICLE{Laitinen2015,
       author = {{Laitinen}, T. and {Huttunen-Heikinmaa}, K. and {Valtonen}, E. and {Dalla}, S.},
        title = "{Correcting for Interplanetary Scattering in Velocity Dispersion analysis of Solar Energetic Particles}",
      journal = {\apj},
         year = 2015,
        month = jun,
       volume = {806},
       number = {1},
          eid = {114},
        pages = {114},
          doi = {10.1088/0004-637X/806/1/114},
archivePrefix = {arXiv},
       eprint = {1504.06166},
 primaryClass = {astro-ph.SR},
       adsurl = {https://ui.adsabs.harvard.edu/abs/2015ApJ...806..114L}
}

@ARTICLE{Lionello2009,
       author = {{Lionello}, Roberto and {Linker}, Jon A. and {Miki{\'c}}, Zoran},
        title = "{Multispectral Emission of the Sun During the First Whole Sun Month: Magnetohydrodynamic Simulations}",
      journal = {\apj},
         year = 2009,
        month = jan,
       volume = {690},
       number = {1},
        pages = {902-912},
          doi = {10.1088/0004-637X/690/1/902},
       adsurl = {https://ui.adsabs.harvard.edu/abs/2009ApJ...690..902L}
}

@ARTICLE{McComas2006,
       author = {{McComas}, D.~J. and {Schwadron}, N.~A.},
        title = "{An explanation of the Voyager paradox: Particle acceleration at a blunt termination shock}",
      journal = {\grl},
         year = 2006,
        month = feb,
       volume = {33},
       number = {4},
          eid = {L04102},
        pages = {L04102},
          doi = {10.1029/2005GL025437},
       adsurl = {https://ui.adsabs.harvard.edu/abs/2006GeoRL..33.4102M}
}

@ARTICLE{McComas2016,
       author = {{McComas}, D.~J. and {Alexander}, N. and {Angold}, N. and {Bale}, S. and {Beebe}, C. and {Birdwell}, B. and {Boyle}, M. and {Burgum}, J.~M. and {Burnham}, J.~A. and {Christian}, E.~R. and {Cook}, W.~R. and {Cooper}, S.~A. and {Cummings}, A.~C. and {Davis}, A.~J. and {Desai}, M.~I. and {Dickinson}, J. and {Dirks}, G. and {Do}, D.~H. and {Fox}, N. and {Giacalone}, J. and {Gold}, R.~E. and {Gurnee}, R.~S. and {Hayes}, J.~R. and {Hill}, M.~E. and {Kasper}, J.~C. and {Kecman}, B. and {Klemic}, J. and {Krimigis}, S.~M. and {Labrador}, A.~W. and {Layman}, R.~S. and {Leske}, R.~A. and {Livi}, S. and {Matthaeus}, W.~H. and {McNutt}, R.~L. and {Mewaldt}, R.~A. and {Mitchell}, D.~G. and {Nelson}, K.~S. and {Parker}, C. and {Rankin}, J.~S. and {Roelof}, E.~C. and {Schwadron}, N.~A. and {Seifert}, H. and {Shuman}, S. and {Stokes}, M.~R. and {Stone}, E.~C. and {Vandegriff}, J.~D. and {Velli}, M. and {von Rosenvinge}, T.~T. and {Weidner}, S.~E. and {Wiedenbeck}, M.~E. and {Wilson}, P.},
        title = "{Integrated Science Investigation of the Sun (ISIS): Design of the Energetic Particle Investigation}",
      journal = {\ssr},
         year = 2016,
        month = dec,
       volume = {204},
       number = {1-4},
        pages = {187-256},
          doi = {10.1007/s11214-014-0059-1},
       adsurl = {https://ui.adsabs.harvard.edu/abs/2016SSRv..204..187M}
}

@ARTICLE{Muller2020,
       author = {{M{\"u}ller}, D. and {St. Cyr}, O.~C. and {Zouganelis}, I. and {Gilbert}, H.~R. and {Marsden}, R. and {Nieves-Chinchilla}, T. and {Antonucci}, E. and {Auch{\`e}re}, F. and {Berghmans}, D. and {Horbury}, T.~S. and {Howard}, R.~A. and {Krucker}, S. and {Maksimovic}, M. and {Owen}, C.~J. and {Rochus}, P. and {Rodriguez-Pacheco}, J. and {Romoli}, M. and {Solanki}, S.~K. and {Bruno}, R. and {Carlsson}, M. and {Fludra}, A. and {Harra}, L. and {Hassler}, D.~M. and {Livi}, S. and {Louarn}, P. and {Peter}, H. and {Sch{\"u}hle}, U. and {Teriaca}, L. and {del Toro Iniesta}, J.~C. and {Wimmer-Schweingruber}, R.~F. and {Marsch}, E. and {Velli}, M. and {De Groof}, A. and {Walsh}, A. and {Williams}, D.},
        title = "{The Solar Orbiter mission. Science overview}",
      journal = {\aap},
         year = 2020,
        month = oct,
       volume = {642},
          eid = {A1},
        pages = {A1},
          doi = {10.1051/0004-6361/202038467},
archivePrefix = {arXiv},
       eprint = {2009.00861},
 primaryClass = {astro-ph.SR},
       adsurl = {https://ui.adsabs.harvard.edu/abs/2020A&A...642A...1M}
}

@BOOK{Reames2017,
  author    = {Reames, D.~V.},
  title     = {{Solar energetic particles}},
  publisher = {Springer-Verlag},
  address   = {Berlin, Germany},
  year      = {2017}
}

@ARTICLE{Reames2023,
       author = {{Reames}, Donald V.},
        title = "{Review and outlook of solar energetic particle measurements on multispacecraft missions}",
      journal = {Frontiers in Astronomy and Space Sciences},
         year = 2023,
        month = aug,
       volume = {10},
          eid = {1254266},
        pages = {1254266},
          doi = {10.3389/fspas.2023.1254266},
archivePrefix = {arXiv},
       eprint = {2307.04182},
 primaryClass = {astro-ph.SR},
       adsurl = {https://ui.adsabs.harvard.edu/abs/2023FrASS..1054266R}
}

@ARTICLE{Riley2011,
       author = {{Riley}, P. and {Lionello}, R. and {Linker}, J.~A. and {Mikic}, Z. and {Luhmann}, J. and {Wijaya}, J.},
        title = "{Global MHD Modeling of the Solar Corona and Inner Heliosphere for the Whole Heliosphere Interval}",
      journal = {\solphys},
         year = 2011,
        month = dec,
       volume = {274},
       number = {1-2},
        pages = {361-377},
          doi = {10.1007/s11207-010-9698-x},
       adsurl = {https://ui.adsabs.harvard.edu/abs/2011SoPh..274..361R}
}

@ARTICLE{Rodriguez-Garcia2023,
       author = {{Rodr{\'\i}guez-Garc{\'\i}a}, L. and {Balmaceda}, L.~A. and {G{\'o}mez-Herrero}, R. and {Kouloumvakos}, A. and {Dresing}, N. and {Lario}, D. and {Zouganelis}, I. and {Fedeli}, A. and {Espinosa Lara}, F. and {Cernuda}, I. and {Ho}, G.~C. and {Wimmer-Schweingruber}, R.~F. and {Rodr{\'\i}guez-Pacheco}, J.},
        title = "{Solar activity relations in energetic electron events measured by the MESSENGER mission}",
      journal = {\aap},
         year = 2023,
        month = jun,
       volume = {674},
          eid = {A145},
        pages = {A145},
          doi = {10.1051/0004-6361/202245604},
archivePrefix = {arXiv},
       eprint = {2212.01592},
 primaryClass = {astro-ph.SR},
       adsurl = {https://ui.adsabs.harvard.edu/abs/2023A&A...674A.145R}
}

@ARTICLE{Rodriguez-Pacheco2020,
       author = {{Rodr{\'\i}guez-Pacheco}, J. and {Wimmer-Schweingruber}, R.~F. and
         {Mason}, G.~M. and {Ho}, G.~C. and {S{\'a}nchez-Prieto}, S. and
         {Prieto}, M. and {Mart{\'\i}n}, C. and {Seifert}, H. and
         {Andrews}, G.~B. and {Kulkarni}, S.~R. and {Panitzsch}, L. and
         {Boden}, S. and {B{\"o}ttcher}, S.~I. and {Cernuda}, I. and
         {Elftmann}, R. and {Espinosa Lara}, F. and {G{\'o}mez-Herrero}, R. and
         {Terasa}, C. and {Almena}, J. and {Begley}, S. and {B{\"o}hm}, E. and
         {Blanco}, J.~J. and {Boogaerts}, W. and {Carrasco}, A. and
         {Castillo}, R. and {da Silva Fari{\~n}a}, A. and
         {de Manuel Gonz{\'a}lez}, V. and {Drews}, C. and {Dupont}, A.~R. and
         {Eldrum}, S. and {Gordillo}, C. and {Guti{\'e}rrez}, O. and
         {Haggerty}, D.~K. and {Hayes}, J.~R. and {Heber}, B. and {Hill}, M.~E. and
         {J{\"u}ngling}, M. and {Kerem}, S. and {Knierim}, V. and
         {K{\"o}hler}, J. and {Kolbe}, S. and {Kulemzin}, A. and {Lario}, D. and
         {Lees}, W.~J. and {Liang}, S. and {Mart{\'\i}nez Hell{\'\i}n}, A. and
         {Meziat}, D. and {Montalvo}, A. and {Nelson}, K.~S. and {Parra}, P. and
         {Paspirgilis}, R. and {Ravanbakhsh}, A. and {Richards}, M. and
         {Rodr{\'\i}guez-Polo}, O. and {Russu}, A. and {S{\'a}nchez}, I. and
         {Schlemm}, C.~E. and {Schuster}, B. and {Seimetz}, L. and
         {Steinhagen}, J. and {Tammen}, J. and {Tyagi}, K. and {Varela}, T. and
         {Yedla}, M. and {Yu}, J. and {Agueda}, N. and {Aran}, A. and
         {Horbury}, T.~S. and {Klecker}, B. and {Klein}, K. -L. and
         {Kontar}, E. and {Krucker}, S. and {Maksimovic}, M. and {Maland
        raki}, O. and {Owen}, C.~J. and {Pacheco}, D. and {Sanahuja}, B. and
         {Vainio}, R. and {Connell}, J.~J. and {Dalla}, S. and {Dr{\"o}ge}, W. and
         {Gevin}, O. and {Gopalswamy}, N. and {Kartavykh}, Y.~Y. and
         {Kudela}, K. and {Limousin}, O. and {Makela}, P. and {Mann}, G. and
         {{\"O}nel}, H. and {Posner}, A. and {Ryan}, J.~M. and {Soucek}, J. and
         {Hofmeister}, S. and {Vilmer}, N. and {Walsh}, A.~P. and {Wang}, L. and
         {Wiedenbeck}, M.~E. and {Wirth}, K. and {Zong}, Q.},
        title = "{The Energetic Particle Detector. Energetic particle instrument suite for the Solar Orbiter mission}",
      journal = {\aap},
         year = 2020,
        month = oct,
       volume = {642},
          eid = {A7},
        pages = {A7},
          doi = {10.1051/0004-6361/201935287},
       adsurl = {https://ui.adsabs.harvard.edu/abs/2020A&A...642A...7R}
}

@ARTICLE{Rouillard2016,
       author = {{Rouillard}, A.~P. and {Plotnikov}, I. and {Pinto}, R.~F. and {Tirole}, M. and {Lavarra}, M. and {Zucca}, P. and {Vainio}, R. and {Tylka}, A.~J. and {Vourlidas}, A. and {De Rosa}, M.~L. and {Linker}, J. and {Warmuth}, A. and {Mann}, G. and {Cohen}, C.~M.~S. and {Mewaldt}, R.~A.},
        title = "{Deriving the Properties of Coronal Pressure Fronts in 3D: Application to the 2012 May 17 Ground Level Enhancement}",
      journal = {\apj},
         year = 2016,
        month = dec,
       volume = {833},
       number = {1},
          eid = {45},
        pages = {45},
          doi = {10.3847/1538-4357/833/1/45},
archivePrefix = {arXiv},
       eprint = {1605.05208},
 primaryClass = {astro-ph.SR},
       adsurl = {https://ui.adsabs.harvard.edu/abs/2016ApJ...833...45R}
}

@ARTICLE{Sarris1976,
       author = {{Sarris}, E.~T. and {Krimigis}, S.~M. and {Armstrong}, T.~P.},
        title = "{Observations of magnetospheric bursts of high-energy protons and electrons at {\ensuremath{\sim}}35 R$_{E}$ with Imp 7}",
      journal = {\jgr},
         year = 1976,
        month = may,
       volume = {81},
       number = {13},
        pages = {2341},
          doi = {10.1029/JA081i013p02341},
       adsurl = {https://ui.adsabs.harvard.edu/abs/1976JGR....81.2341S}
}

@ARTICLE{Sarris1987,
       author = {{Sarris}, E.~T. and {Anagnostopoulos}, G.~C. and {Krimigis}, S.~M.},
        title = "{Simultaneous measurements of energetic ion (>=50 keV) and electron (>=220 keV) activity upstream of earth's bow shock and inside the plasma sheet: Magnetospheric source for the November 3 and December 3, 1977 upstream events}",
      journal = {\jgr},
         year = 1987,
        month = nov,
       volume = {92},
       number = {A11},
        pages = {12083-12096},
          doi = {10.1029/JA092iA11p12083},
       adsurl = {https://ui.adsabs.harvard.edu/abs/1987JGR....9212083S}
}

@ARTICLE{SunPy2020,
       author = {{SunPy Community} and {Barnes}, Will T. and {Bobra}, Monica G. and {Christe}, Steven D. and {Freij}, Nabil and {Hayes}, Laura A. and {Ireland}, Jack and {Mumford}, Stuart and {Perez-Suarez}, David and {Ryan}, Daniel F. and {Shih}, Albert Y. and {Chanda}, Prateek and {Glogowski}, Kolja and {Hewett}, Russell and {Hughitt}, V. Keith and {Hill}, Andrew and {Hiware}, Kaustubh and {Inglis}, Andrew and {Kirk}, Michael S.~F. and {Konge}, Sudarshan and {Mason}, James Paul and {Maloney}, Shane Anthony and {Murray}, Sophie A. and {Panda}, Asish and {Park}, Jongyeob and {Pereira}, Tiago M.~D. and {Reardon}, Kevin and {Savage}, Sabrina and {Sip{\H{o}}cz}, Brigitta M. and {Stansby}, David and {Jain}, Yash and {Taylor}, Garrison and {Yadav}, Tannmay and {Rajul} and {Dang}, Trung Kien},
        title = "{The SunPy Project: Open Source Development and Status of the Version 1.0 Core Package}",
      journal = {\apj},
         year = 2020,
        month = feb,
       volume = {890},
       number = {1},
          eid = {68},
        pages = {68},
          doi = {10.3847/1538-4357/ab4f7a},
       adsurl = {https://ui.adsabs.harvard.edu/abs/2020ApJ...890...68S}
}

@INPROCEEDINGS{Vainio2018,
       author = {{Vainio}, Rami and {Afanasiev}, Alexandr},
        title = "{Particle Acceleration Mechanisms}",
    booktitle = {Solar Particle Radiation Storms Forecasting and Analysis},
         year = 2018,
       editor = {{Malandraki}, Olga E. and {Crosby}, Norma B.},
       series = {Astrophysics and Space Science Library},
       volume = {444},
        month = jan,
        pages = {45-61},
          doi = {10.1007/978-3-319-60051-2_3},
       adsurl = {https://ui.adsabs.harvard.edu/abs/2018ASSL..444...45V}
}

@ARTICLE{Vainio2013,
       author = {{Vainio}, Rami and {Valtonen}, Eino and {Heber}, Bernd and {Malandraki}, Olga E. and {Papaioannou}, Athanasios and {Klein}, Karl-Ludwig and {Afanasiev}, Alexander and {Agueda}, Neus and {Aurass}, Henry and {Battarbee}, Markus and {Braune}, Stephan and {Dr{\"o}ge}, Wolfgang and {Ganse}, Urs and {Hamadache}, Clarisse and {Heynderickx}, Daniel and {Huttunen-Heikinmaa}, Kalle and {Kiener}, J{\"u}rgen and {Kilian}, Patrick and {Kopp}, Andreas and {Kouloumvakos}, Athanasios and {Maisala}, Sami and {Mishev}, Alexander and {Miteva}, Rositsa and {Nindos}, Alexander and {Oittinen}, Tero and {Raukunen}, Osku and {Riihonen}, Esa and {Rodr{\'\i}guez-Gas{\'e}n}, Rosa and {Saloniemi}, Oskari and {Sanahuja}, Blai and {Scherer}, Renate and {Spanier}, Felix and {Tatischeff}, Vincent and {Tziotziou}, Kostas and {Usoskin}, Ilya G. and {Vilmer}, Nicole},
        title = "{The first SEPServer event catalogue \raisebox{-0.5ex}\textasciitilde68-MeV solar proton events observed at 1 AU in 1996-2010}",
      journal = {Journal of Space Weather and Space Climate},
         year = 2013,
        month = mar,
       volume = {3},
          eid = {A12},
        pages = {A12},
          doi = {10.1051/swsc/2013030},
       adsurl = {https://ui.adsabs.harvard.edu/abs/2013JSWSC...3A..12V}
}

@INPROCEEDINGS{Wiedenbeck2017,
       author = {{Wiedenbeck}, M.~E. and {Angold}, N.~G. and {Birdwell}, B. and {Burnham}, J.~A. and {Christian}, E.~R. and {Cohen}, C.~M.~S. and {Cook}, W.~R. and {Cummings}, A.~C. and {Davis}, A.~D. and {Dirks}, G. and {Do}, D.~H. and {Everett}, D.~T. and {Goodwin}, P.~A. and {Hanley}, J.~J. and {Hernandez}, L. and {Kecman}, B. and {Klemic}, J. and {Labrador}, A.~W. and {Leske}, R.~A. and {Lopez}, S. and {Link}, J.~T. and {McComas}, D.~J. and {Mewaldt}, R.~A. and {Miyasaka}, H. and {Nahory}, B.~W. and {Rankin}, J.~S. and {Riggans}, G. and {Rodriguez}, B. and {Rusert}, M.~D. and {Shuman}, S.~A. and {Simms}, K.~M. and {Stone}, E.~C. and {von Rosenvinge}, T.~T. and {Weidner}, S.~E. and {White}, M.~L.},
        title = "{Capabilities and Performance of the High-Energy Energetic-Particles Instrument for the Parker Solar Probe Mission}",
    booktitle = {35th International Cosmic Ray Conference (ICRC2017)},
         year = 2017,
       series = {International Cosmic Ray Conference},
       volume = {301},
        month = jul,
          eid = {16},
        pages = {16},
          doi = {10.22323/1.301.0016},
       adsurl = {https://ui.adsabs.harvard.edu/abs/2017ICRC...35...16W}
}

@ARTICLE{Wilson2025,
       author = {{Wilson}, Lynn B. and {Mitchell}, J. Grant and {Szabo}, Adam and {Jebaraj}, Immanuel C. and {Stevens}, Michael L. and {Malaspina}, David M. and {Berland}, Grant D. and {Kouloumvakos}, Athanasios and {Bale}, Stuart D. and {Livi}, Roberto and {Halekas}, Jasper S. and {Cohen}, Christina M.~S.},
        title = "{Large-amplitude Whistler Precursors and >MeV Particles Observed at a Weak Interplanetary Shock by Parker Solar Probe}",
      journal = {\apj},
         year = 2025,
        month = jul,
       volume = {987},
       number = {1},
          eid = {31},
        pages = {31},
          doi = {10.3847/1538-4357/add6a8},
       adsurl = {https://ui.adsabs.harvard.edu/abs/2025ApJ...987...31W}
}

@ARTICLE{Wimmer2021,
       author = {{Wimmer-Schweingruber}, R.~F. and {Janitzek}, N.~P. and {Pacheco}, D. and {Cernuda}, I. and {Espinosa Lara}, F. and {G{\'o}mez-Herrero}, R. and {Mason}, G.~M. and {Allen}, R.~C. and {Xu}, Z.~G. and {Carcaboso}, F. and {Kollhoff}, A. and {K{\"u}hl}, P. and {Freiherr von Forstner}, J.~L. and {Berger}, L. and {Rodriguez-Pacheco}, J. and {Ho}, G.~C. and {Andrews}, G.~B. and {Angelini}, V. and {Aran}, A. and {Boden}, S. and {B{\"o}ttcher}, S.~I. and {Carrasco}, A. and {Dresing}, N. and {Eldrum}, S. and {Elftmann}, R. and {Evans}, V. and {Gevin}, O. and {Hayes}, J. and {Heber}, B. and {Horbury}, T.~S. and {Kulkarni}, S.~R. and {Lario}, D. and {Lees}, W.~J. and {Limousin}, O. and {Malandraki}, O.~E. and {Mart{\'\i}n}, C. and {O'Brien}, H. and {Prieto Mateo}, M. and {Ravanbakhsh}, A. and {Rodriguez-Polo}, O. and {S{\'a}nchez Prieto}, S. and {Schlemm}, C.~E. and {Seifert}, H. and {Terasa}, J.~C. and {Tyagi}, K. and {Vainio}, R. and {Walsh}, A. and {Yedla}, M.~K.},
        title = "{First year of energetic particle measurements in the inner heliosphere with Solar Orbiter's Energetic Particle Detector}",
      journal = {\aap},
         year = 2021,
        month = dec,
       volume = {656},
          eid = {A22},
        pages = {A22},
          doi = {10.1051/0004-6361/202140940},
archivePrefix = {arXiv},
       eprint = {2108.02020},
 primaryClass = {astro-ph.SR},
       adsurl = {https://ui.adsabs.harvard.edu/abs/2021A&A...656A..22W}
}

@INPROCEEDINGS{Wuelser2004,
       author = {{Wuelser}, Jean-Pierre and {Lemen}, James R. and {Tarbell}, Theodore D. and {Wolfson}, C.~J. and {Cannon}, Joseph C. and {Carpenter}, Brock A. and {Duncan}, Dexter W. and {Gradwohl}, Glenn S. and {Meyer}, Syndie B. and {Moore}, Augustus S. and {Navarro}, Rosemarie L. and {Pearson}, J.~D. and {Rossi}, George R. and {Springer}, Larry A. and {Howard}, Russell A. and {Moses}, John D. and {Newmark}, Jeffrey S. and {Delaboudiniere}, Jean-Pierre and {Artzner}, Guy E. and {Auchere}, Frederic and {Bougnet}, Marie and {Bouyries}, Philippe and {Bridou}, Francoise and {Clotaire}, Jean-Yves and {Colas}, Gerard and {Delmotte}, Franck and {Jerome}, Arnaud and {Lamare}, Michel and {Mercier}, Raymond and {Mullot}, Michel and {Ravet}, Marie-Francoise and {Song}, Xueyan and {Bothmer}, Volker and {Deutsch}, Werner},
        title = "{EUVI: the STEREO-SECCHI extreme ultraviolet imager}",
    booktitle = {Telescopes and Instrumentation for Solar Astrophysics},
         year = 2004,
       editor = {{Fineschi}, Silvano and {Gummin}, Mark A.},
       series = {Society of Photo-Optical Instrumentation Engineers (SPIE) Conference Series},
       volume = {5171},
        month = feb,
        pages = {111-122},
          doi = {10.1117/12.506877},
       adsurl = {https://ui.adsabs.harvard.edu/abs/2004SPIE.5171..111W}
}

@ARTICLE{Xiaohang2025,
       author = {{Chen}, Xiaohang and {Zhao}, Lulu and {Giacalone}, Joe and {Sachdeva}, Nishtha and {Sokolov}, Igor and {Toth}, Gabor and {Cohen}, Christina and {Lario}, David and {Guo}, Fan and {Kouloumvakos}, Athanasios and {Gombosi}, Tamas and {Huang}, Zhenguang and {Manchester}, Ward and {van der Holst}, Bart and {Liu}, Weihao and {McComas}, David and {Hill}, Matthew and {Ho}, George},
        title = "{Evidence of Time-Dependent Diffusive Shock Acceleration in the 2022 September 5 Solar Energetic Particle Event}",
      journal = {arXiv e-prints},
         year = 2025,
        month = jun,
          eid = {arXiv:2506.20322},
        pages = {arXiv:2506.20322},
          doi = {10.48550/arXiv.2506.20322},
archivePrefix = {arXiv},
       eprint = {2506.20322},
 primaryClass = {astro-ph.SR},
       adsurl = {https://ui.adsabs.harvard.edu/abs/2025arXiv250620322C}
}

@ARTICLE{Xu2026arXiv,
       author = {{Xu}, Zigong and {Cohen}, C.~M.~S. and {Leske}, R.~A. and {Muro}, G.~D. and {Cummings}, A.~C. and {Romeo}, O.~M. and {Lario}, D. and {McComas}, D.~J. and {Cuesta}, M.~E. and {Pak}, S. and {Khoo}, L.~Y. and {Farooki}, H.~A. and {Shen}, M.~M. and {Kasapis}, S. and {Christian}, E.~R. and {Mitchell}, D.~G. and {McNutt}, R.~L. and {Kouloumvakos}, A. and {Mitchell}, J. Grant and {Berland}, G.~D. and {Schwadron}, N.~A. and {Wiedenbeck}, M.~E. and {Stevens}, M.~L. and {Allen}, R.~C.},
        title = "{Parker Solar Probe observations of solar energetic particle (SEP) events with inverse velocity arrival (IVA) features}",
      journal = {arXiv e-prints},
         year = 2026,
        month = feb,
          eid = {arXiv:2602.12475},
        pages = {arXiv:2602.12475},
          doi = {10.48550/arXiv.2602.12475},
archivePrefix = {arXiv},
       eprint = {2602.12475},
 primaryClass = {astro-ph.SR},
       adsurl = {https://ui.adsabs.harvard.edu/abs/2026arXiv260212475X}
}

@article{Yuncong2025,
    author = {Li, Yuncong and Guo, Jingnan and Pacheco, Daniel and Wang, Yuming and Temmer, Manuela and Ding, Zheyi and Wimmer-Schweingruber, Robert F},
    title = {The delayed arrival of faster solar energetic particles as a probe into the shock acceleration process},
    journal = {National Science Review},
    volume = {12},
    number = {10},
    pages = {nwaf348},
    year = {2025},
    month = {08},
    issn = {2095-5138},
    doi = {10.1093/nsr/nwaf348},
    url = {https://doi.org/10.1093/nsr/nwaf348},
    eprint = {https://academic.oup.com/nsr/article-pdf/12/10/nwaf348/64138238/nwaf348.pdf},
}
\bibliographystyle{aasjournalv7}

\appendix

\section{List of IVA-SEP events and shock properties}

Table~\ref{tab:event_list} lists the IVA-SEP events identified in the survey. Column (2) lists the date of the event, and columns (3) and (4) the associated flare and the flare start peak time respectively. Columns (5) and (6) list the flare location (longitude and latitude) in Stonyhurst coordinates) from EUV data. Column (7) list the spacecraft that observed the event and column (8) the heliocentric distance of the spacecraft. Column (9) and (10) list the transition energy (E$_t$) and its uncertainty that both determined from a visual inspection of the energy spectrograms. Column (11) list the CME width from CDAW and (12) notes for which of the IVA-SEPs a shock fitting is performed.

\begin{table}[!ht]
\centering
\caption{List of IVA-SEP events.}
\label{tab:event_list}
\begin{tabular}{cccccccccccc}
\hline
Event & Date & \multicolumn{2}{c}{Flare} & \multicolumn{2}{c}{Location$^{(b)}$}  & s/c & r$_{s/c}$ & E$_t$ & E$_t^u$  & CME & Shock \\
\# & & Class$^{(a)}$ & Peak & Lon.\,[$^\circ$] & Lat.\,[$^\circ$] & & [AU] & [MeV] & [MeV] & Width\,[$^\circ$]  & Fit. \\
(1) & (2) & (3) & (4) & (5) & (6) & (7) & (8) & (9) & (10) & (11) & (12) \\ [1ex]
\hline

1	&	2022-06-07	&	C1.1(S)	&	2022-06-07T03:27	&	163	&	20	&	SolO	&	0.96	&	0.9	&	[0.5 - 1]	&	240	&	Y	\\
2	&	2022-06-26	&	C1.8(S)	&	2022-06-26T03:19	&	147	&	-28	&	SolO	&	1.01	&	1.5	&	[0.9 - 3]	&	360	&	Y	\\
3	&	2022-07-23	&	C1.4(S)	&	2022-07-23T20:32	&	176	&	-28	&	SolO	&	0.99	&	8.0	&	[5 - 12]	&	360	&	Y	\\
4	&	2023-01-20	&	$<$C2	&	2023-01-20T14$^{(c)}$	&	45	&	-36	&	SolO	&	0.95	&	1.0	&	[0.8 - 3]	&	176	&	Y	\\
5	&	2023-03-13	&	(N)	&	(N)	&	(N)	&	(N)	&	SolO	&	0.61	&	1.5	&	[0.7 - 3]	&	360	&	Y	\\
6	&	2023-05-02	&	C1.1(S)	&	2023-05-02T04:37	&	$\sim$150	&	(N)	&	SolO	&	0.52	&	2.0	&	[1 - 5]	&	360	&	Y	\\
7	&	2023-11-02	&	C5.0(G)	&	2023-11-02T05:10	&	-23	&	22	&	SolO	&	0.57	&	0.6	&	[0.4 - 1.5]	&	125	&	Y	\\
8	&	2023-11-09	&	C2.6(G)	&	2023-11-09T11:13	&	16	&	-17	&	SolO	&	0.66	&	2.0	&	[1 - 3]	&	360	&	Y	\\
9	&	2023-12-24	&	$<$C1	&	…	&	31	&	46	&	SolO	&	0.94	&	2.0	&	[1 - 4]	&	194	&	Y	\\
10	&	2023-12-31	&	C5.6(G)	&	2023-12-31T12:54	&	41	&	27	&	SolO	&	0.95	&	2.5	&	[2 - 4]	&	146	&	Y	\\
11	&	2024-01-21	&	C5.3(G)	&	2024-01-21T00:40	&	-42	&	24	&	SolO	&	0.79	&	(U)	&		&	360	&	Y	\\
12	&	2024-03-23	&	X1.1(G)	&	2024-03-23T01:24	&	-8	&	23	&	SolO	&	0.39	&	15.0	&	[10 - 20]	&	360	&	Y	\\
13	&	2024-10-24	&	X3.3(G)	&	2024-10-24T03:47	&	-89	&	-18	&	SolO	&	0.54	&	(U)	&		&	360	&	Y	\\
14	&	2024-11-04	&	M3.8(G)	&	2024-11-04T01:35	&	-44	&	-6	&	SolO	&	0.67	&	0.6	&	[0.4 - 1]	&	360	&	Y	\\
15	&	2024-12-29	&	X1.1(G)	&	2024-12-29T07:15	&	-30	&	-16	&	SolO	&	0.95	&	(U)	&		&	360	&	Y	\\
16	&	2022-09-05	&	M8.9(S)	&	2022-09-05T16:17	&	175	&	-30	&	PSP	&	0.08	&	2.0	&	[1 - 3]	&	360	&	Y	\\
17	&	2024-06-24	&	C1.1(G)	&	2024-06-24T04:44	&	88	&	-27	&	PSP	&	0.29	&	(U)	&		&	239	&	Y	\\
18	&	2024-09-22	&	M3.5(G)	&	2024-09-22T21:26	&	-64	&	-22	&	PSP	&	0.35	&	(U)	&		&	360	&	Y	\\
19	&	2024-09-29	&	\textbf{C5.0(G)}	&	\textbf{2024-09-29T06:12}	&	35	&	20	&	PSP	&	0.09	&	(U)	&		&	130	&	Y	\\

\hline

1*	&	2023-04-21	&	B9.5(G)	&	2023-04-21T05:58	&	$\sim$90	&	$\sim$9	&	SolO	&	0.37	&	0.2	&	[0.1 - 0.4]	&	49	&	N	\\
2*	&	2023-08-08	&	(U)	&	(U)	&	(U)	&	(U)	&	SolO	&	0.88	&	(U)	&		&		&	N	\\
3*	&	2024-05-08	&	(U)	&	(U)	&	(U)	&	(U)	&	SolO	&	0.67	&	0.6	&	[0.4 - 1]	&		&	N	\\
4*	&	2024-06-01	&	M2.9(S)	&	2024-06-01T18:46	&	-168	&	-8	&	SolO	&	0.86	&	(U)	&		&	360	&	N	\\
5*	&	2024-09-18	&	C2.2(S)	&	2024-09-18T12:25	&	-89	&	-17	&	SolO	&	0.4	&	1.5	&	[1 - 5]	&	161	&	N	\\
6*	&	2025-03-17	&	C2.2(G)	&	2025-03-17T11:00	&	32	&	11	&	SolO	&	0.42	&	1.0	&	[0.5 - 2]	&	90	&	N	\\
7*	&	2024-10-03	&	(U)	&	(U)	&	(U)	&	(U)	&	PSP	&	0.17	&	(U)	&		&	&	N	\\

\hline
\end{tabular}
\vspace{0.5em}

\begin{minipage}{1\textwidth}
\small
$^{(a)}$: Flare class from GOES (G) (or STIX (S) for occulted flares) . \\
$^{(b)}$: Location of the eruption from EUV observations, in Stonyhurst coordinates. \\
$^{(c)}$: Approximate flare peak time from EUV observations. \\
$^{*}$: Events that left out of the analysis (see main text for details). \\
(N): Col. 3: No flare observations available; Col. 5 \& 6: No EUV observations available to determine flare location. \\
(U): Col. 3--6: Multiple eruptions or lack of observations lead to uncertain event association. Col. 9: Uncertain E$_t$ determination. \\

\end{minipage}
\end{table}

Table~\ref{tab:shock_param} lists the shock parameters determined from the 3D modeling. Columns (1) and (2) list the number and date of the events. Columns (3) and (4) list the location of the shock apex (longitude and latitude, respectively) in Stonyhurst coordinates. Columns (5) and (6) list the shock speed determined and the apex and at the COBPOINTs. Columns (7) and (8) list the mean and maximum $M_{\text{fms}}$, and Column (9) lists the mean $\lambda$-angle, at the COBPOINTs.

\begin{table*}[!ht]
\centering
\caption{Shock Parameters from 3D modeling.}
\label{tab:shock_param}
\begin{tabular}{ccccccccc}
\hline
Event & Date & \multicolumn{2}{c}{Apex direction$^{(a)}$} & \multicolumn{2}{c}{Speed} & \multicolumn{2}{c}{$M_{\text{fms}}$} & $\lambda$-angle \\
 & & Lon. [$^\circ$] & Lat. [$^\circ$] & Apex$^{(b)}$ [km/s] & FLs [km/s] & mean & max & [$^\circ$] \\
(1) & (2) & (3) & (4) & (5) & (6) & (7) & (8) & (9) \\ [1ex]
\hline

1	&	2022-06-07	&	168.0	&	19.8	&	1154	&	709	&	2.6	&	3.5	&	99.5	\\
2	&	2022-06-26	&	158.8	&	-23.6	&	1192	&	893	&	4.3	&	5.2	&	96.0	\\
3	&	2022-07-23	&	-169.8	&	-9.6	&	1007	&	1124	&	5.0	&	7.3	&	36.1	\\
4	&	2023-01-20	&	46.0	&	-34.5	&	846	&	731	&	0.7	&	0.8	&	57.0	\\
5	&	2023-03-13	&	-128.9	&	18.5	&	1964	&	1387	&	2.4	&	3.6	&	143.9	\\
6	&	2023-05-02	&	154.5	&	-26.2	&	1293	&	1050	&	6.9	&	7.3	&	72.1	\\
7	&	2023-11-02	&	-20.4	&	33.8	&	421	&	501	&	1.9	&	3.2	&	75.1	\\
8	&	2023-11-09	&	14.0	&	10.2	&	807	&	786	&	3.1	&	3.4	&	9.9	\\
9	&	2023-12-24	&	1.3	&	50.4	&	492	&	594	&	0.8	&	2.2	&	82.3	\\
10	&	2023-12-31	&	54.9	&	24.3	&	480	&	649	&	1.0	&	2.3	&	51.1	\\
11	&	2024-01-21	&	-44.7	&	13.4	&	1066	&	659	&	1.3	&	2.0	&	123.1	\\
12	&	2024-03-23	&	2.3	&	28.4	&	1424	&	1353	&	6.1	&	7.6	&	66.7	\\
13	&	2024-10-24	&	-70.2	&	-28.7	&	2491	&	1225	&	2.0	&	3.1	&	129.5	\\
14	&	2024-11-04	&	-43.4	&	7.7	&	1306	&	563	&	1.0	&	1.3	&	123.1	\\
15	&	2024-12-29	&	-33.2	&	-26.2	&	1091	&	578	&	0.9	&	1.3	&	126.5	\\
16	&	2022-09-05	&	170.3	&	-40.5	&	2405	&	1833	&	3.3	&	4.7	&	92.4	\\
17	&	2024-06-24	&	88.1	&	-30.6	&	486	&	580	&	0.3	&	0.5	&	66.3	\\
18	&	2024-09-22	&	-60.0	&	-38.0	&	1751	&	714	&	0.9	&	1.8	&	128.8	\\
19	&	2024-09-29	&	60.3	&	-21.7	&	…	&	820	&	3.1	&	3.6	&	38.6	\\

\hline
\end{tabular}
\vspace{0.5em}

\hspace*{10em}
\begin{minipage}{1\textwidth}
\small
$^{(a)}$: Mean direction of shock's apex propagation in Stonyhurst coordinates. \\
$^{(b)}$: The apex speed determined at 5~R$_\Sun$. For event \#19 the reconstruction starts above 5~R$_\Sun$.
\end{minipage}
\end{table*}

\newpage
\section{Details on Shock 3D Reconstruction and Kinematics} \label{append:3DRec}

\begin{figure}[!h]
    \centering
    \includegraphics[width=0.75\textwidth]{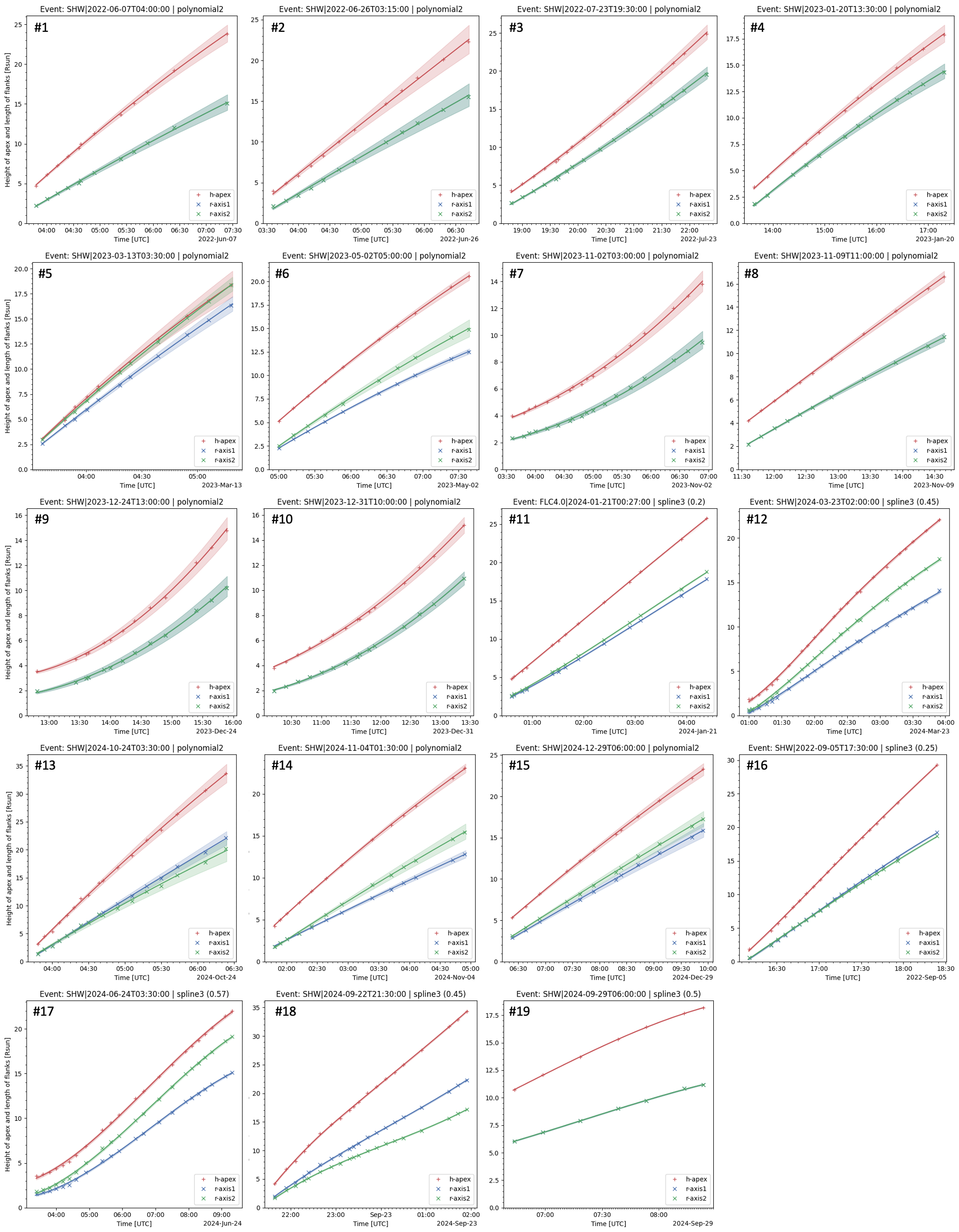}
    \caption{Time profiles of the fittings of the reconstructed 3D shock for all events analyzed in this study. Each panel (\#1--\#19) shows the heliocentric distance of the shock apex (red crosses) and the lengths of the two semi-axes of the fitted ellipsoid (blue and green crosses) derived from the 3D reconstruction. Solid curves indicate the best-fit functions (second-order polynomial or cubic spline, as noted in each panel), and the shaded bands denote the corresponding fit uncertainty. Times are given in UTC, and heliocentric distances are in solar radii (R$_\odot$).}
    \label{fig:Shock_Kinematics_append}
\end{figure}

For the reconstruction analysis we used PyThea, which is a software package written in Python language that can be used to reconstruct the structure of CMEs and shock waves in 3D \citep{Kouloumvakos2022_Pythea}. This reconstruction process involves the fitting, at different times, of an ellipsoid, to achieve the best visual agreement of the model to the multi‑viewpoint remote sensing observations. The cadence of the remote sensing data used, is one minute for AIA and 5 minutes for STEREO-A EUVI data, and ranges from 5 to 30 minutes for the coronagraphic data in white light depending on the instrument (typically $\sim$15 min. for COR2 ($\sim$5~min. in high-cadence campaigns), $\sim$12~min. for C2 and $\sim$30~min. for C3). In many cases, the shock fittings start from distances $\geq$2.5~R$_\sun$ (e.g., at the LASCO-C2 field-of-view and above) because EUV waves were identified in only a few events. We further discuss implications of this in Section~\ref{sec:Discussion}. In Figure~\ref{fig:Shock_Kinematics_append}, we show for each event the time profiles of the height of the apex and the radius of the two semi-axes from the 3D fitting of the ellipsoid to the images, and the fittings with their uncertainty.

Overall the 3D Reconstruction can significantly reduce projection effects by combining observations from multiple viewpoints, providing a more accurate determination of the shock location and its kinematics. However, for many events in this study, the geometric constraints were limited because STEREO‑A and near‑Earth spacecraft were not ideally positioned --separated by an angular distance of 27~degrees trailing Earth's orbit in mid-2022 to 16~degrees ahead in mid-2024-- to provide comprehensive information of the shocks' three‑dimensional structure. Another issue for the 3D reconstruction is that, in some cases, it was impossible to trace unambiguously the shock fronts in every direction, so we had to rely on indirect signatures of the location of the disturbance such as the deflection of streamers. This worked well for most cases, but for a few events the final uncertainty may be greater than typically expected \citep[e.g.,][that showed an expected uncertainty of $\sim$10\%]{Kwon2014}.

\newpage
\section{Changes in magnetic connections} \label{append:model}

\begin{figure}[!h]
    \centering
    \includegraphics[width=0.9\textwidth]{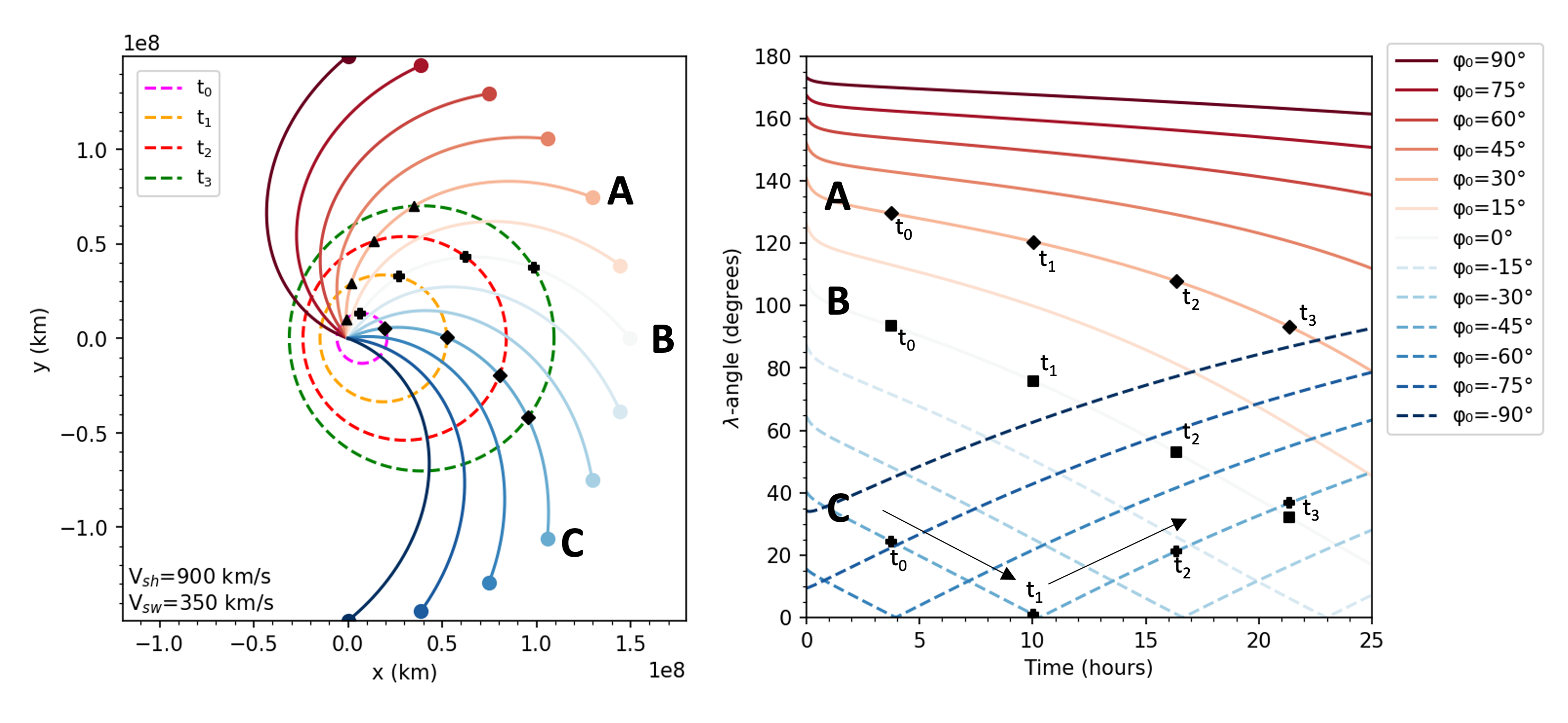}
    \caption{Temporal evolution of the central separation angle, $\lambda$, for observers located at different heliolongitudes with respect to the shock apex, derived from a 2D circular shock model assuming a shock speed of 900~km/s. Top panel shows the model geometry of the expanding shock fronts at four times (colored circular arcs), the Parker-spiral magnetic field lines connecting each observer to the front assuming a solar wind speed of 350~km/s, and observers distributed in longitude around the Sun. The shock apex propagates toward $\phi = 0^\circ$ and three representative observers are highlighted: B, located along the apex direction, and A and C, located west and east of the apex, respectively. The bottom panel shows the evolution of $\lambda$-angle at the COBPOINTs for each observer as a function of time. Black symbols trace the evolution of $\lambda$-angle at the COBPOINTs of the three highlighted observers (A–C) at the four times indicated in the top panel.}
    \label{fig:Toy_model}
\end{figure}

In Figure~\ref{fig:Toy_model}, we present an example of the temporal evolution of the central separation angle for multiple observers located at different longitudes relative to the shock apex ($\phi=0^\circ$), based on a simple 2D shock model. In this example, observers initially connected to the western flanks (i.e., A and B in Figure~\ref{fig:Toy_model}) experience a connectivity shift toward the apex within a few hours (see bottom panel of Figure~\ref{fig:Toy_model}), while those on the eastern side (i.e., C in Figure~\ref{fig:Toy_model})) progressively connect to regions farther from the apex. Since shock regions near the apex are generally stronger than those at the flanks, western observers first sample weaker shock regions and subsequently stronger ones, a sequence that possibly favors the development of IVA formation. Conversely, eastern observers experience the opposite effect. 

\end{document}